\documentclass[sigconf]{setting/acmart}
\usepackage{booktabs}   
\usepackage{subcaption} 
\usepackage{enumitem}
\usepackage{algorithm}
\usepackage{algpseudocode}
\usepackage{multicol}
\usepackage{siunitx} 
\usepackage{etoolbox}
\usepackage{soul}
\usepackage{cancel}
\usepackage[normalem]{ulem}

\usepackage[font=small,skip=0pt]{caption}

\usepackage{amsmath,tabularx}

\makeatletter
\newcommand{\multiline}[1]{%
    \begin{tabularx}{\dimexpr\linewidth-\ALG@thistlm}[t]{@{}X@{}}
    #1
    \end{tabularx}
}

\usepackage[compact]{titlesec}

\renewcommand\thesubsubsection{\Alph{subsection}.\arabic{subsubsection}}
\renewcommand\theparagraph{\underline{\Alph{subsection}.\arabic{subsubsection}.\alph{paragraph}}}

\titleformat{\subsubsection}[runin]
{\normalfont\normalsize\itshape}{\scalebox{.8}{$\blacksquare$} \thesubsubsection)}{.6em}{}[\ \ ]
\titlespacing*{\subsubsection}
{0pt}{.3ex plus .2ex minus .2ex}{.3ex plus .2ex minus .2ex}

\titleformat{\paragraph}[runin]
{\normalfont\normalsize\itshape}{\theparagraph}{.5em}{}[\ \ ]
\titlespacing*{\paragraph}
{0pt}{.2ex plus 1ex minus .2ex}{.2ex plus .2ex}

\titlespacing*{\section}{0pt}{0.1\baselineskip}{0.2\baselineskip}
\algtext*{EndFor}
\algtext*{EndIf}
\algtext*{EndProcedure}

\begin{document}

\pdfcompresslevel=9
\pdfminorversion=5
\pdfobjcompresslevel=2
\setcopyright{none}
\renewcommand\footnotetextcopyrightpermission[1]{}
\settopmatter{printccs=false,printacmref=false}
\settopmatter{printacmref=false} 
\pagestyle{plain}

\title{TurboFNO: High-Performance Fourier Neural Operator with Fused FFT-GEMM-iFFT on GPU\vspace{-0mm}}

\author{Shixun Wu}
\affiliation{%
  \institution{University of California, Riverside}
  \city{Riverside}
  \state{CA}
  \country{USA}}
\email{swu264@ucr.edu}

\author{Yujia Zhai}
\affiliation{%
  \institution{University of California, Riverside}
  \city{Riverside}
  \state{CA}
  \country{USA}}
\email{yzhai015@ucr.edu}

\author{Huangliang Dai}
\affiliation{%
  \institution{University of California, Riverside}
  \city{Riverside}
  \state{CA}
  \country{USA}}
\email{hdai022@ucr.edu}

\author{Hairui Zhao}
\affiliation{%
  \institution{University of California, Riverside}
  \city{Riverside}
  \state{CA}
  \country{USA}}
\email{hairuiz@ucr.edu}

\author{Yue Zhu}
\affiliation{%
  \institution{University of California, Riverside}
  \city{Riverside}
  \state{CA}
  \country{USA}}
\email{yzhu303@ucr.edu}

\author{Haiyang Hu}
\affiliation{%
  \institution{University of California, Riverside}
  \city{Riverside}
  \state{CA}
  \country{USA}}
\email{hhu064@ucr.edu}





\author{Zizhong Chen}
\affiliation{%
  \institution{University of California, Riverside}
  \city{Riverside}
  \state{CA}
  \country{USA}}
\email{chen@cs.ucr.edu}

\begin{abstract}
  Fourier Neural Operators (FNO) are widely used for learning partial differential equation solution operators. However, FNO lacks architecture-aware optimizations,with its Fourier layers executing FFT, filtering, GEMM, zero padding, and iFFT as separate stages, incurring multiple kernel launches and significant global memory traffic. We propose TurboFNO, the first fully fused FFT-GEMM-iFFT GPU kernel with built-in FFT optimizations. We first develop FFT and GEMM kernels from scratch, achieving performance comparable to or faster than the closed-source SOTA cuBLAS and cuFFT. Additionally, our FFT kernel integrates a built-in high-frequency truncation, input zero-padding, and pruning feature
 to avoid additional memory copy kernels. To fuse the FFT and GEMM workloads, we propose an FFT variant in which a single thread block iterates over the hidden dimension, aligning with the $k$-loop in GEMM. Additionally, we design two shared memory swizzling patterns to achieve 100\% memory bank utilization when forwarding FFT output to GEMM and enabling the iFFT to retrieve GEMM results directly from shared memory.Experimental result on an NVIDIA A100 GPU shows TurboFNO outperforms PyTorch, cuBLAS, and cuFFT by up to 150\%.
\end{abstract}
\maketitle
\section{Introduction}
\begin{figure}
    \centering
    \includegraphics[width=\linewidth]{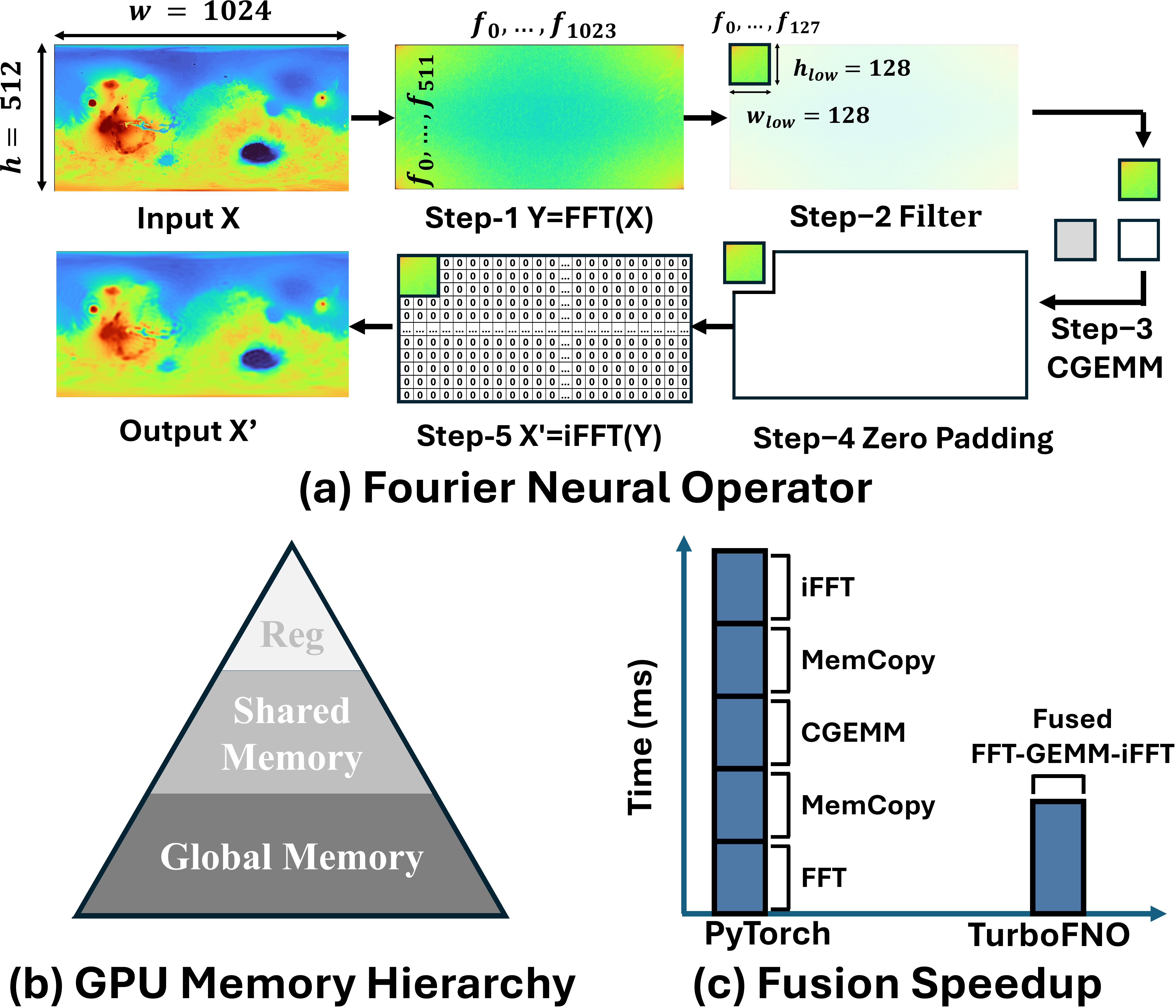}
    \caption{Fourier Neural Operator and TurboFNO}
    \label{fig:FNO_illustration}
\end{figure}

Fourier Neural Operator (FNO) \cite{li2020fourier} has become a powerful learning framework for solving partial differential equations (PDEs) by leveraging the spectral representation of functions. Through the use of the Fast Fourier Transform (FFT), FNO projects input functions to the frequency domain, applies learned transformations via matrix multiplications, and returns to the spatial domain using the inverse FFT (iFFT). This approach achieves strong generalization and has been widely applied in scientific domains such as fluid dynamics, weather forecasting, and quantum mechanics \cite{pathak2022fourcastnet,li2021neural}. Figure~\ref{fig:FNO_illustration}(a) illustrates the overall workflow of the Fourier Neural Operator (FNO), which transforms input functions into the frequency domain via FFT, applies a learned linear transformation, and returns to the spatial domain using iFFT. 

The FFT → GEMM → iFFT computational pattern is not unique to Fourier Neural Operators; rather, it is a fundamental motif in a wide range of scientific computing applications \cite{bertsch2002density,goedecker1999linear}. For example, in quantum chemistry and materials science, solving the time-dependent Schrödinger equation or simulating real-time time-dependent density functional theory (RT-TDDFT) often involves repeated FFT-based transformations followed by dense matrix multiplications \cite{chen2010classical}. These simulations rely on FFT to move between spatial and spectral domains, GEMM to apply potential or interaction operators in Fourier space, and iFFT to return to the spatial domain for further computation. Similar workflows appear in electronic structure calculations \cite{kotliar2006electronic}, turbulence modeling \cite{wilcox1998turbulence,scotti2002turbulence}, signal processing \cite{johnston2023curriculum,johnston2023downlink}, scientific data compression \cite{liu2024high,jian2024cliz,huang2023exploring,liu2023stationary,liu2024cusz} and signal propagation in electromagnetics \cite{liao2006taylor}. In many of these applications, the FFT-GEMM-iFFT routine is applied to large, high-dimensional data grids with tight memory and latency constraints, making performance optimization critical \cite{plimpton2017ffts,murashima2019coupling,habib2013hacc}.

Despite its ubiquity, the FFT~$\rightarrow$~GEMM~$\rightarrow$~iFFT pattern has received limited attention in terms of GPU kernel fusion and memory layout optimization. Existing domain-specific libraries---such as \textit{Quantum ESPRESSO} \cite{giannozzi2009quantum}, \textit{Octopus} \cite{castro2006octopus}, and \textit{CP2K} \cite{hutter2014cp2k}---typically invoke separate calls to FFT and BLAS routines, leading to redundant memory transfers and suboptimal GPU utilization. Even in the context of Fourier Neural Operators, the most widely used implementations (e.g., PyTorch-based \cite{paszke2019pytorch}) rely on separate invocations of cuFFT \cite{nvidia_cufft}, cuBLAS \cite{nvidia_cublas}, and custom frequency filter kernels. These implementations lack architecture-aware optimizations, resulting in inefficient execution on modern GPUs. In particular, current implementations suffer from three key limitations: 1) \textbf{Partial frequency utilization:} In the standard FFT~$\rightarrow$~GEMM~$\rightarrow$~iFFT workflow, the GEMM operation is applied only to selected low-frequency components. Supporting this pattern requires additional memory copy kernels to extract and insert the relevant frequencies, incurring nontrivial overhead. 2) \textbf{Lack of native frequency filtering in cuFFT:} cuFFT does not natively support frequency filtering. Moreover, its closed-source design makes integrating customized filtering logic difficult or impractical. 3) \textbf{Excessive memory transactions:} Because cuFFT and cuBLAS are implemented as independent black-box libraries, intermediate results must be stored in global memory between FFT, GEMM, and iFFT stages. This leads to additional memory transactions and limits opportunities for data reuse and kernel fusion.

As neural operator models such as FNO and its variants are increasingly applied to large-scale scientific simulations, there is a growing demand for high-performance, scalable implementation. Recent work in this area focuses on optimizing deep learning workloads through a variety of techniques, including operator-level tuning, memory layout transformation, communication-aware scheduling, and hardware-specific code generation. Frameworks such as TVM \cite{chen2018tvm}, Triton \cite{tillet2019triton}, and TensorRT \cite{nvidia_tensorrt} have demonstrated the importance of low-level performance optimization in enabling real-time or large-scale deployment \cite{zhaoarraypipe,zhao2024visage,li2023explsched,li2024interference}. However, these tools are largely general-purpose and are not tailored to the unique compute pattern of Fourier Neural Operators, which involve a structured FFT → GEMM → iFFT pipeline. In the context of FNO, existing infrastructure rarely exploits opportunities for data reuse and fusion across spectral transforms and linear layers, resulting in suboptimal memory traffic and performance.

Despite the fundamental importance of FFT and GEMM in scientific computing and AI workloads, to the best of our knowledge, no prior work has explored how to fuse FFT and GEMM kernels. Kernel fusion is a key optimization technique in modern deep learning systems to reduce kernel launch overhead, minimize global memory traffic, and improve overall hardware utilization \cite{wu2023anatomy,wu2023ft,wu2025turbofft,wu2024ft,wu2024dgro,dai2025ft,zukswarm}. A number of recent high-impact systems employ kernel fusion to accelerate specific models or operations. For example, ByteTransformer \cite{zhai2023bytetransformer} and NVIDIA’s FasterTransformer \cite{chelba2020faster} apply kernel fusion to optimize large transformer models, reducing latency in transformer blocks by fusing layer normalization, matrix multiplication, and activation functions. FlashAttention \cite{dao2022flashattention} fuses attention computation with softmax and dropout in a memory-efficient manner to accelerate transformer inference and training. Similarly, DeepSpeed \cite{rasley2020deepspeed}, and Megatron-LM \cite{shoeybi2019megatron} incorporate various kernel fusion strategies to improve training throughput on large-scale models.  Existing efforts in kernel fusion predominantly focus on FFT–convolution~\cite{fu2023flashfftconv} and FFT–stencil~\cite{han2025flashfftstencil} pipelines, where the dataflow structures are more naturally aligned. In contrast, fusing FFT and GEMM presents unique challenges due to their mismatched data access patterns and memory layouts, and remains an open area of research.

Fusing FFT and GEMM kernels is inherently challenging due to their fundamentally mismatched dataflows. For example, a typical FFT kernel computes a workload of size $2^N \times 8$ per thread block, which does not naturally align with GEMM’s blocked matrix structure. However, in the context of Fourier Neural Operators (FNO), we observe a unique opportunity for alignment arising from its use of frequency-domain filtering. In particular, commonly used truncation thresholds in FNO—such as 32 or 64—effectively reshape the FFT output to match the size of GEMM input tiles. As shown in Figure~\ref{fig:Dataflow}, the dataflow in FNO’s spectral convolution (Figure~\ref{fig:FNO_illustration}, Step 3) reveals that the GEMM operation along the hidden dimension shares a natural interface with the preceding FFT and succeeding iFFT stages. Building upon this insight, we propose \textbf{TurboFNO}, a fully fused kernel design that tightly integrates FFT, CGEMM, and iFFT into a unified workflow. TurboFNO incorporates several performance optimizations, including built-in FFT truncation, zero padding, and pruning, as well as a carefully designed alignment of FFT and GEMM dataflows. To further ensure high efficiency, we introduce warp-level \textbf{thread swizzling strategies} to eliminate shared memory bank conflicts during inter-stage data movement. These combined innovations enable TurboFNO to outperform the baseline PyTorch implementation across a wide range of configurations. More specifically, our contributions include:
\begin{itemize}[leftmargin=*]
  \item \textbf{TurboFNO}: We propose \textit{TurboFNO}, the first fully fused FFT-GEMM-iFFT kernel tailored for Fourier Neural Operators. The entire FFT, truncation, CGEMM, zero-padding, and iFFT pipeline is integrated into a single GPU kernel, avoiding intermediate memory transactions and kernel launches.
  
  \item \textbf{Custom FFT and GEMM kernels}: To enable fusion, we develop FFT and GEMM kernels from scratch, achieving performance comparable to or better than the closed-source cuFFT and cuBLAS.
  
  \item \textbf{Built-in frequency truncation, zero-padding, and pruning}: Our FFT kernel supports native frequency filtering and input zero-padding, removing the need for additional truncation and padding kernels and significantly reducing memory overhead. A pruning strategy is used to avoid redundant computation.
  
  \item \textbf{Fused FFT-GEMM-iFFT}: To fuse the FFT and GEMM workload, we propose a FFT variant that make one threadblock iterate over the hidden dimension to fuse with the k-loop in GEMM. Besides, we design two shared memory swizzling patterns to achieve 100\% memory bank utilization when forwarding FFT output to GEMM, and iFFT retrieves GEMM results from shared memory.

  \item \textbf{Significant performance improvement}: Experimental results show that TurboFNO outperforms PyTorch’s cuFFT + cuBLAS-based implementation by up to 150\%, demonstrating the value of architecture-aware co-design in scientific neural operators.
\end{itemize}

\section{Background}
\label{sec:background}


In this section, we provide the necessary background on the architecture and workflow of Fourier Neural Operators (FNO), and prior work on FFT-related kernel fusion. We also highlight how TurboFNO extends these foundations with architecture-aware designs.

\subsection{Fourier Neural Operators}

Figure~\ref{fig:FNO_illustration}(a) illustrates the typical FNO pipeline, which applies a forward FFT to input feature maps, performs spectral filtering, and then reconstructs the output via an inverse FFT. Notably, FNO often requires truncation (Step 2) and zero-padding (Step 4) in the frequency domain. In existing frameworks such as PyTorch, these operations are performed by separate memory-copy kernels because cuFFT does not natively support input/output trimming or padding. This results in additional kernel launches and global memory traffic. Moreover, as shown in Figure~\ref{fig:Dataflow}, the input to each Fourier layer has the shape \texttt{[Batch, HiddenDim, DimX, DimY]}, where the hidden dimension typically ranges from 64 to 128. After applying the first FFT, the feature map is truncated in the frequency domain, reducing its shape to \texttt{[Batch, HiddenDim, dimX, dimY]} with $dimX < DimX$ and $dimY < DimY$ based on the truncation ratio. Subsequently, a complex-valued linear layer (i.e., CGEMM) is applied along the \texttt{HiddenDim} axis. This matrix multiplication yields an output with the same shape as the truncated input. To reconstruct the full spatial resolution, the output is then zero-padded back to \texttt{DimX} $\times$ \texttt{DimY}, followed by the second FFT stage or inverse FFT, depending on whether it's a forward or backward spectral transformation. This dataflow reinforces the benefit of aligning memory layouts across FFT, GEMM, and iFFT stages to eliminate redundant memory transfers and maximize shared memory reuse. 

Figures~\ref{fig:FNO_illustration}(b) and (c) highlight the core motivation behind TurboFNO. Due to the nature of FNO’s frequency filter—which discards high-frequency components—existing PyTorch-based implementations perform substantial \textbf{redundant global memory reads/writes and unnecessary FFT computations}. These inefficiencies result in significant performance bottlenecks, particularly when applied to large-scale problems. TurboFNO addresses these challenges by introducing a high-performance, fully fused FFT and GEMM kernel, along with a carefully aligned dataflow that enables efficient end-to-end kernel fusion. This design dramatically reduces memory traffic and computation overhead, leading to significant speedups over baseline implementations.

\subsection{Built-in FFT Optimization}

Existing state-of-the-art FFT libraries, such as cuFFT \cite{cufft}, and VkFFT  \cite{tolmachev2023vkfft} failed to support the data truncation feature. In contrast, TurboFNO implements a high-performance custom FFT kernel that supports \textbf{built-in truncation and zero-padding}, by modifying the global memory input/output layout directly. Based on this foundation, we observe that truncation discards a large portion of the frequency spectrum—often more than half—which leads to unnecessary butterfly computations in standard FFT execution. TurboFNO therefore introduces a GPU-side \textbf{FFT pruning strategy} that eliminates redundant operations in these truncated frequency bands, significantly reducing computation and memory bandwidth.

\subsection{FFT-CGEMM Fusion via Dataflow Alignment}

Beyond FFT optimization alone, TurboFNO proposes a novel \textbf{fusion strategy} between FFT, CGEMM, and iFFT by aligning their dataflows. In traditional 2D FFTs, both 1D-FFT stages are applied along the spatial dimensions (e.g., height and width). However, in FNO, the spectral weights (i.e., CGEMM) are applied across the channel or hidden dimension. This motivates us to decouple the two FFT stages: we retain the first 1D FFT along the spatial axis, but \textbf{reinterpret the second FFT stage along the hidden dimension}. 

This reinterpretation enables a batched-FFT layout that mirrors the \texttt{k}-loop of GEMM, allowing each thread block to load data along the same direction as operand $A$ in matrix multiplication. Furthermore, FFT outputs can be written directly into shared memory, forming the input tile for CGEMM. After CGEMM, the blocked output matrix $C$ can remain in shared memory and be used immediately as input to the inverse FFT, entirely bypassing global memory between stages.

To our knowledge, this workflow is novel. Prior FFT fusion work has primarily focused on FFT–convolution–iFFT chains, where the input and output tensor dimensions remain fixed and fusion is relatively straightforward. In contrast, fusing FFT and CGEMM in FNO requires a deep understanding of both algorithms' dataflow, shared memory layout, and performance constraints—making this a fundamentally more challenging fusion problem.

\begin{figure}
    \centering
    \includegraphics[width=\linewidth]{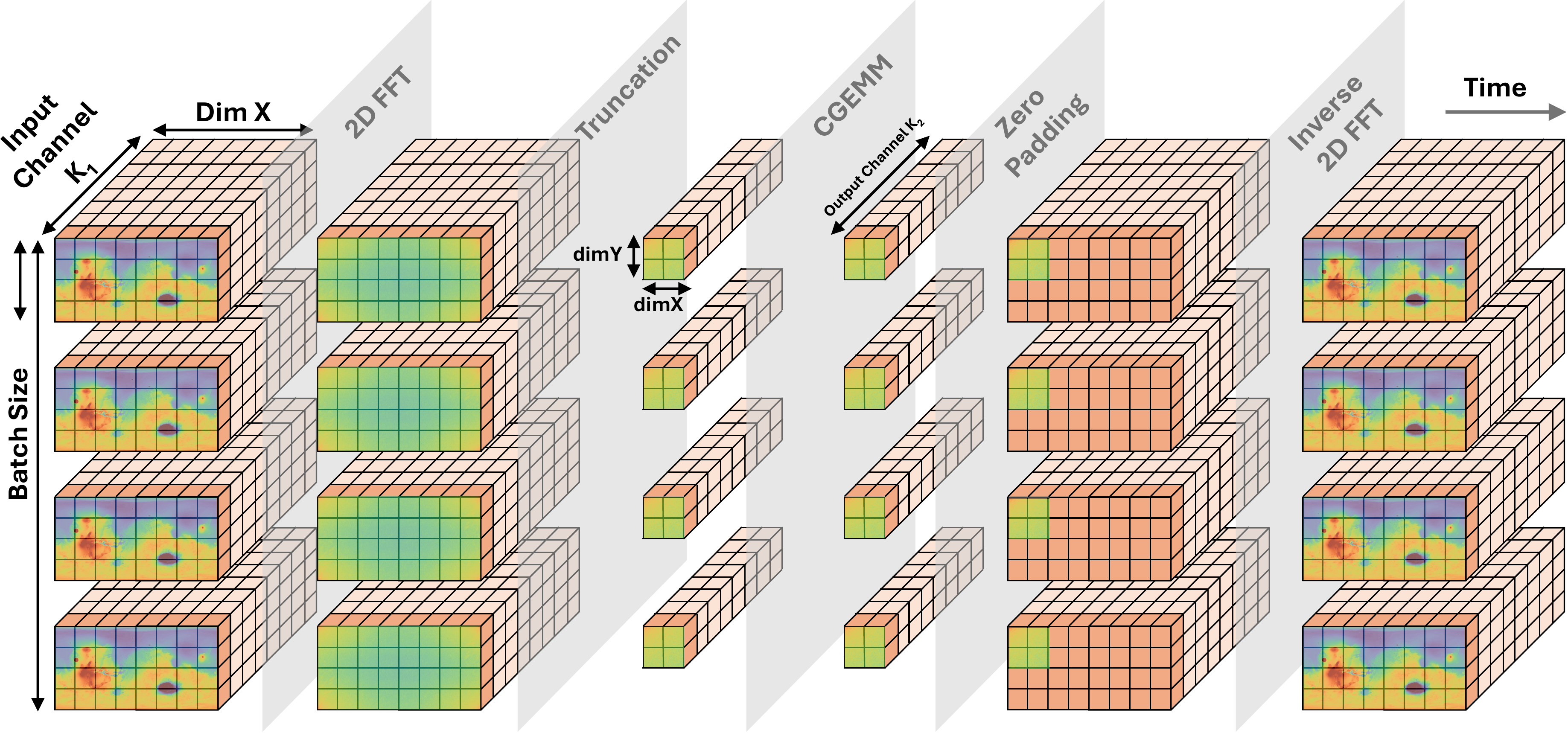}
        \caption{Fourier Nerual Operator Dataflow}
    \label{fig:Dataflow}
\end{figure}
\section{TurboFNO GEMM and FFT Optimization}
To enable kernel fusion and built-in FFT optimizations—including truncation, zero padding, and pruning—a high-performance implementation of both FFT and CGEMM is essential. TurboFNO begins by rethinking the design of these two fundamental components, developing custom GPU kernels that  match state-of-the-art closed-source libraries such as cuFFT and cuBLAS. In the following sections, we first detail how TurboFNO optimizes the CGEMM and FFT kernels to support shared memory dataflow and fusion. We then elaborate on how we integrate FFT-specific optimizations, including pruning of redundant computations and built-in support for truncated and zero-padded signal transformations.

\begin{figure}
    \centering
    \includegraphics[width=\linewidth]{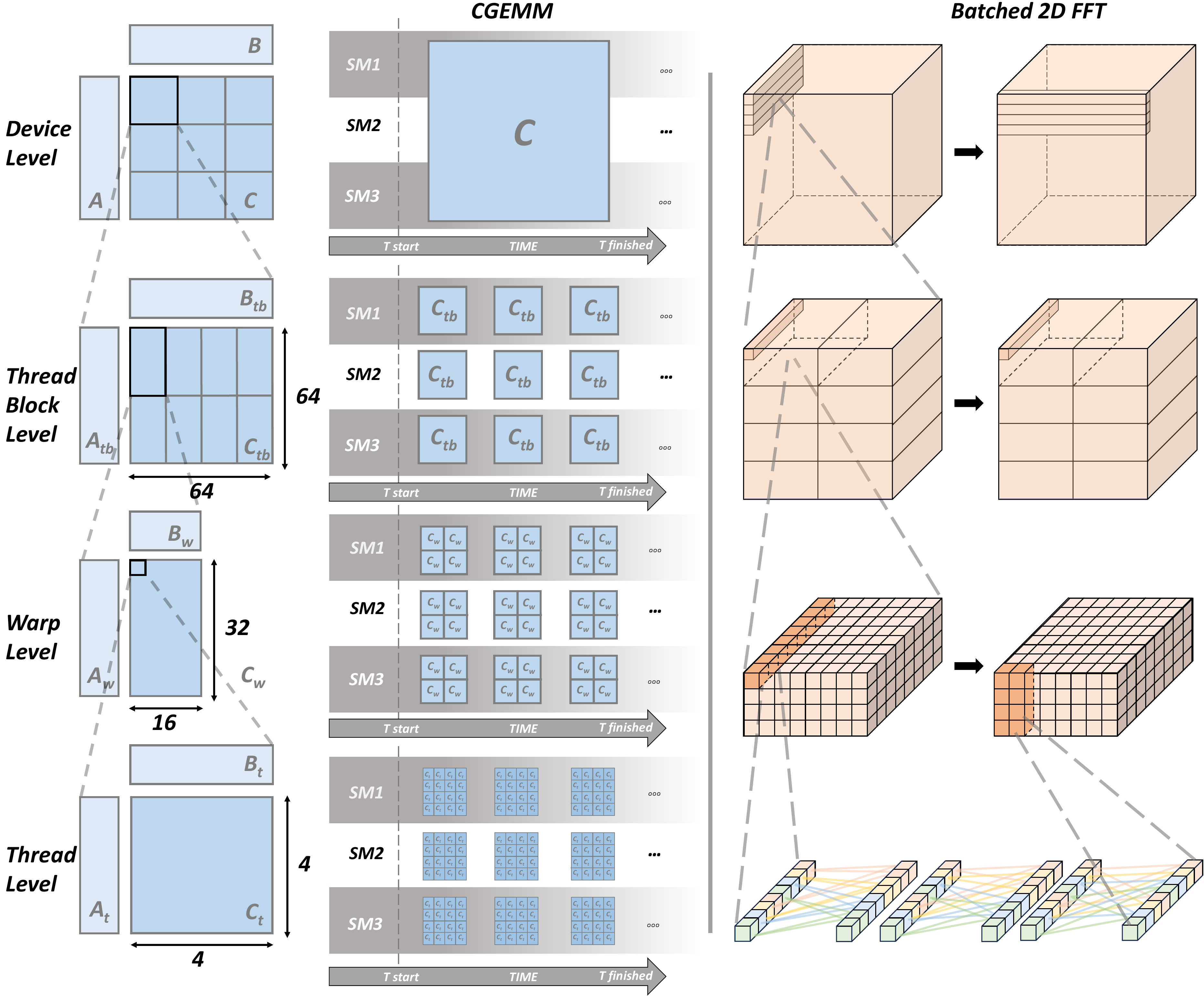}
        \caption{CGEMM and FFT}
    \label{fig:Dataflow}
\end{figure}
\subsection{GEMM in TurboFNO}
We now discuss the implementation of the CGEMM kernel in TurboFNO. Since the baseline PyTorch implementation uses single precision, our focus is on a high-performance \textbf{CUDA-core-based CGEMM} with a blocked design and double shared memory buffering. Although our work focuses on CGEMM (complex-valued GEMM), it is designed to be compatible with later fusion stages. Importantly, the fusion of FFT and GEMM in TurboFNO is achieved by treating the FFT output as the $A$ block in CGEMM. Specifically, FFT reads its input from global memory and writes its output directly into shared memory in a layout that mimics the $A$ operand. This approach does not interfere with the usage of Tensor Cores, though we do not rely on them in this work. 

Another key consideration is the shared memory usage. FFT requires more shared memory than CGEMM due to truncation: in our evaluation, we consider two truncation ratios—25\% and 50\%—where the truncated FFT output occupies only a portion (e.g., $\texttt{dimY} / \texttt{DimY}$) of the original spatial size. Since FFT already introduces synchronization overhead (e.g., \texttt{\_\_syncthreads()} calls during butterfly stages), there is no need to apply double buffering to the $A$ block (i.e., the FFT output). Thus, synchronous copy optimizations (e.g., cooperative global-to-shared memory copy) are unlikely to yield additional benefit in this context. Overall, this work focuses on fusing the FFT and GEMM workflows themselves, rather than on hardware-specific instruction optimizations. Therefore, we exclusively target CUDA cores and rely on general shared memory load/store instructions for compatibility and portability.

Figure~\ref{fig:Dataflow} (left) illustrates the CGEMM workflow of TurboFNO across the GPU execution hierarchy, from the device level down to individual threads. Our implementation adopts a blocked GEMM structure with the following parameter configuration: \texttt{M\_tb} = 64, \texttt{N\_tb} = 64, \texttt{K\_tb} = 8, \texttt{m\_w} = 32, \texttt{n\_w} = 16, and \texttt{m\_thread} = \texttt{n\_thread} = 4. This configuration achieves performance comparable to cuBLAS under large-batch workloads. In the context of FNO, the GEMM operation corresponds to a \texttt{batched CGEMM}, where the effective batch size is the product of the original \texttt{BatchSize} and the spatial extent \texttt{DimX}. The matrix dimensions involved are: $M = \texttt{BatchSize} \times \texttt{DimX} \times \texttt{DimY}$, $N = \texttt{OutputDim}$, and $K = \texttt{HiddenDim}$. Since \texttt{OutputDim} and \texttt{HiddenDim} are typically moderate in size (ranging from a few tens to 256), while $M$ is often very large (e.g., with $\texttt{DimX} = \texttt{DimY} = 64$, we have $M = 4096 \times \texttt{BatchSize}$), the resulting GEMM is a classic case of a \textbf{tall-and-skinny matrix multiplication}. While Figure~\ref{fig:Dataflow} presents one representative configuration, TurboFNO implements a fully \textbf{templated CGEMM kernel} that supports flexible tuning of thread block shapes and loop tiling factors. The templated kernel structure is detailed in Figure~\ref{fig:pseudocode}, enabling us to generalize across diverse problem shapes and maximize GPU utilization across varying workloads.

\subsection{FFT in TurboFNO}
As illustrated on the right side of Figure~\ref{fig:Dataflow}, TurboFNO maps FFT workloads onto the GPU hierarchy with a batched 2D FFT structure. The two FFT stages are performed sequentially: one along the depth axis (into the page) and one along the horizontal axis (rightward). At the \textbf{device level}, each thread block is assigned a batch of four "pencils"—i.e., 1D signals that each require FFT processing. The central idea behind our fusion strategy is to adjust the direction along which these pencils are selected. When we select them along the \textbf{GEMM $k$-loop direction} (i.e., the \texttt{HiddenDim} axis in FNO), the FFT threadblock-level workload can perfectly align with the GEMM operand $A$ block layout, enabling efficient fusion.

However, at the \textbf{warp level}, the memory access pattern becomes a second critical factor in determining fusion efficiency. Even if the thread block selects pencils in a way that aligns with GEMM's dataflow, a mismatch in how FFT output data is organized in threads versus GEMM’s column-major expectation (as shown in Figure~\ref{fig:Dataflow}, left side) can lead to severe shared memory transaction overhead. In such cases, kernel fusion may not provide benefits and can even degrade performance. To support coalesced global memory reads during FFT computation, we adopt the Stockham formulation of FFT, which ensures that each thread reads data in a contiguous pattern. However, writing the results back to shared memory after all butterfly stages requires careful design. As shown in Figure~\ref{fig:Dataflow} (right side, warp level), each thread holds eight FFT output elements in registers, which must be written into shared memory with minimal bank conflict.

Some open-source frameworks such as VkFFT \cite{tolmachev2023vkfft} adopt a strided memory layout strategy in which consecutive threads each write the same-offset data from different pencils. This layout minimizes bank conflict, but—as illustrated in Figure~\ref{fig:shmem_fft_cgemm}(a)—it results in a data layout that is incompatible with GEMM’s column-major expectation. Since each pencil represents a distinct $k$-loop index in GEMM, assigning different $k$-fragments to consecutive threads disrupts shared memory access locality during GEMM, leading to excessive bank conflicts when loading operand $A$. To resolve this, TurboFNO adopts the alternative layout shown in the lower half of Figure~\ref{fig:shmem_fft_cgemm}(a), which reorganizes thread-to-data assignments such that each thread holds consecutive elements from the same pencil, enabling both conflict-free FFT writes and column-major alignment for subsequent CGEMM consumption. We elaborate on this solution and the fusion design between FFT$\rightarrow$GEMM and GEMM$\rightarrow$iFFT in the following section.

\begin{figure}
    \centering
    \includegraphics[width=1\linewidth]{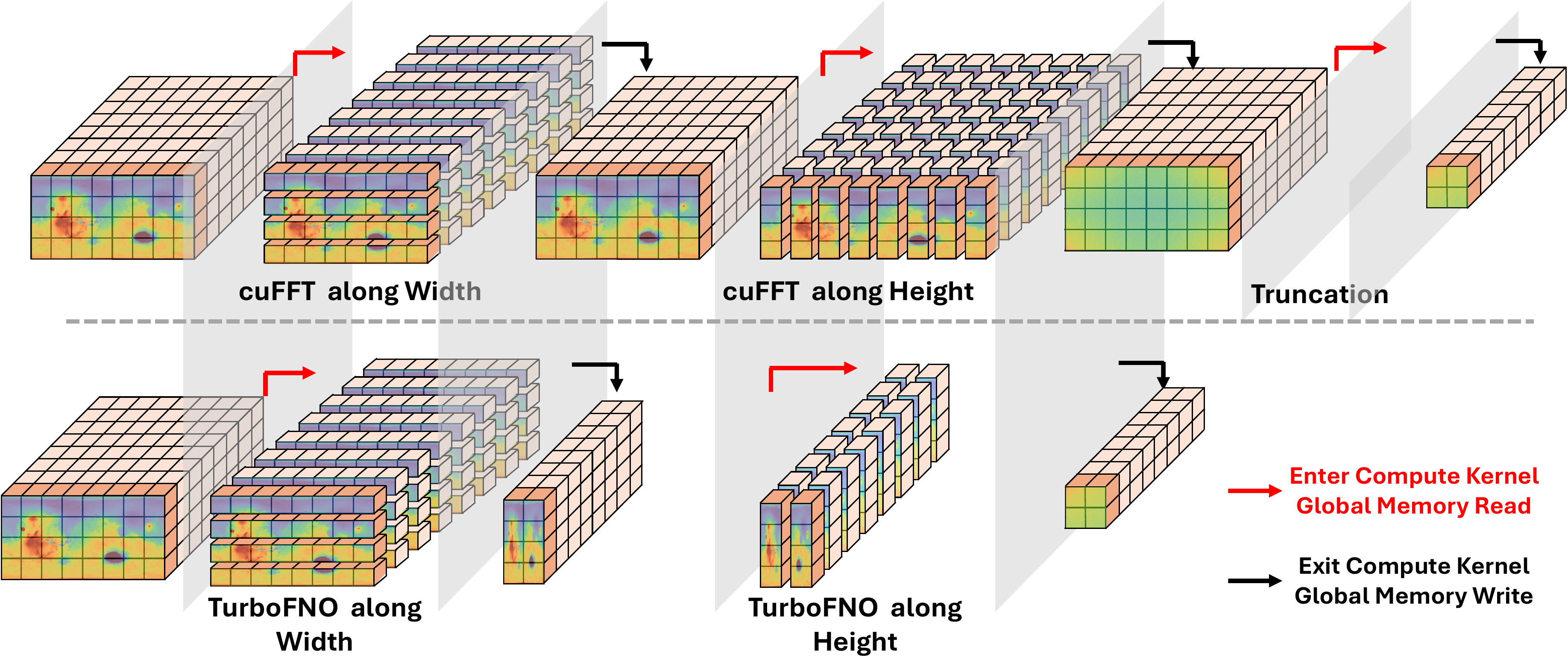}
    \caption{FFT Global Memory}
    \label{fig:trunc_zero_padding}
\end{figure}

\subsection{FFT Optimization}
In addition to shared memory fusion design, Figure~\ref{fig:trunc_zero_padding} compares the FFT workflows of PyTorch and TurboFNO, illustrating TurboFNO’s support for \textbf{built-in truncation and zero padding}. During the first stage of the 2D FFT (along the width dimension, i.e., \texttt{DimX}), TurboFNO only writes the first $dimX / DimX$ fraction of frequency components back to global memory. For example, when $dimX = 64$ and $DimX = 256$, this corresponds to a \textbf{75\% reduction in global memory writes}.

Furthermore, this truncation directly reduces both the \textbf{computation and global memory read of the second FFT stage} (along \texttt{DimY}), since it operates on fewer frequency rows. In the 2D case, the reduction in computational load is quadratic: $(\texttt{dimX}/\texttt{DimX})^2$. As a result, we observe in our experiments that for 2D FNO, kernel fusion built on top of FFT optimization yields only a \textbf{3\%--5\% additional speedup}. In contrast, for 1D FNO where only the second FFT stage is present, TurboFNO's fused FFT-CGEMM-iFFT achieves up to a \textbf{10\% performance improvement}, as the optimization benefits are less masked by the first-stage overhead.
\begin{figure}[h]
    \centering
    \includegraphics[width=1\linewidth]{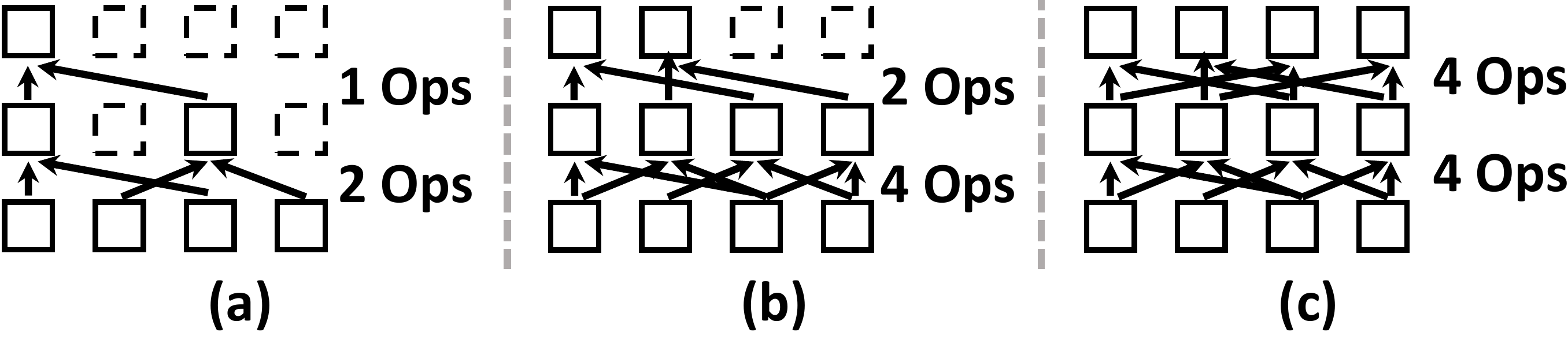}
    \caption{FFT Prune}
    \label{fig:fft_prune}
\end{figure}
In 1D FFT (stage-2), truncation again reduces both the computation and the final global memory write bandwidth by 75\%. As shown in the right-hand side of Figure~\ref{fig:trunc_zero_padding}, the impact of \textbf{zero padding} is complementary to truncation—it similarly reduces the number of required operations and memory accesses, but on the input side.

Additionally, the reduced output or input size allows us to \textbf{prune FFT butterfly stages}. Since only a subset of outputs (e.g., $dimX/DimX$) is needed, many butterfly operations—especially the multiplication phases—can be skipped entirely or replaced by simple additions. This pruning strategy is visualized in Figure~\ref{fig:fft_prune}. In the baseline 4-point FFT (Figure~\ref{fig:fft_prune}c), two stages are required with a total of 8 operations. When truncation ratio is 25\% (Figure~\ref{fig:fft_prune}a), only 3 operations are needed—representing \textbf{37.5\% of the original computation}. At 50\% truncation (Figure~\ref{fig:fft_prune}b), 6 operations are retained—\textbf{75\% of the original workload}. Together, these techniques highlight TurboFNO’s unique capability to \textbf{eliminate redundant computation and global memory traffic} through integrated FFT-layer truncation, zero padding, and butterfly pruning—something not achievable via library-based FFT implementations such as cuFFT.

\section{TurboFNO FFT-CGEMM-iFFT Kernel Fusion}

In this section, we detail the kernel fusion strategies used in TurboFNO, covering three progressive levels of fusion: \textbf{(1) FFT-GEMM fusion}, \textbf{(2) GEMM-iFFT fusion}, and \textbf{(3) full FFT-GEMM-iFFT fusion}. For each stage, we elaborate on how we align the FFT workflow with GEMM, resolve data layout mismatches, and eliminate shared memory bank conflicts. Specifically, we describe how we modify the FFT output layout to match GEMM’s input operand format, and how we apply \textbf{thread swizzling} techniques to ensure conflict-free shared memory access. Furthermore, we show how the inverse FFT (iFFT) is directly integrated as an \textbf{epilogue stage} of CGEMM, bypassing intermediate global memory writes and further improving memory locality. Finally, we present the complete fused FFT-GEMM-iFFT pseudocode, which combines these techniques into a single GPU kernel for maximum performance. The kernel parameters of CGEMM and FFT in TurboFNO is given at Table \ref{tab:kernel_size}.

\subsection{Fused FFT-GEMM}

\begin{figure}
    \centering
    \includegraphics[width=1\linewidth]{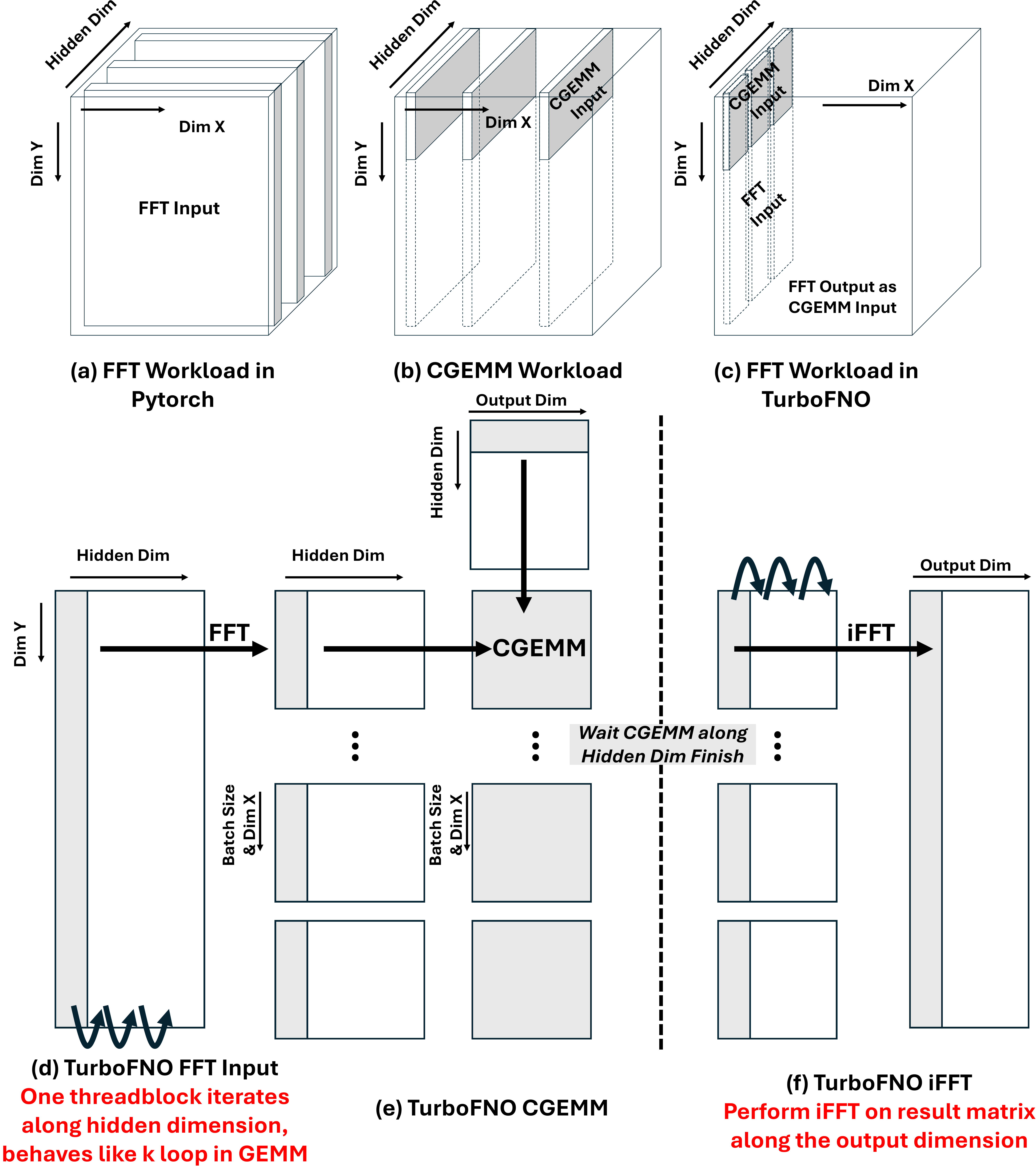}
    \caption{FFT-CGEMM-iFFT Workflow Aignment}
    \label{fig:workflow_alignment}
\end{figure}

Figure~\ref{fig:workflow_alignment} (a)--(c) illustrate how TurboFNO aligns the workflows of FFT and CGEMM to enable kernel fusion. In Figure~\ref{fig:workflow_alignment}(a), we show the default FFT workflow in PyTorch, which performs a 2D FFT along the \texttt{DimX} and \texttt{DimY} axes. However, the subsequent CGEMM operates along the \texttt{HiddenDim} axis, requiring a full writeback of the 2D FFT result to global memory before GEMM can proceed.

We observe that the second 1D FFT stage (along \texttt{DimY}) processes the same spatial region that serves as operand $A$ in CGEMM. The only mismatch lies in the direction of iteration: FFT typically processes data along the \texttt{DimX} axis, whereas CGEMM accesses data in the \texttt{DimY--HiddenDim} plane, iterating along \texttt{HiddenDim}. Thus, to align the workflows, we redesign the second FFT stage to operate along \texttt{HiddenDim} rather than \texttt{DimX}.

As shown in Figure~\ref{fig:workflow_alignment}(c), each thread block fetches a \textbf{slice of input data} along the \texttt{HiddenDim} direction for FFT computation. This slice exactly matches the data block required by CGEMM, shown in Figure~\ref{fig:workflow_alignment}(b). By restructuring the FFT workload in this way, we enable a direct match between the FFT output and CGEMM input. Additionally, we leverage TurboFNO’s built-in truncation capability to reduce the FFT output size before it is forwarded to shared memory. As shown in Figure~\ref{fig:workflow_alignment}(d), the gray region corresponds to the FFT input slice read in (c), and the truncated output is placed directly into shared memory in a format that matches CGEMM’s blocked $A$ layout. The same thread block then performs MAC operations in CGEMM and continues iterating along \texttt{HiddenDim}, fetching the next input slice, performing FFT, and forwarding the result to GEMM—seamlessly integrating into the $k$-loop structure of the GEMM kernel. This alignment of FFT workflow with CGEMM data access not only eliminates redundant memory transfers but also enables shared memory reuse and full fusion without compromising data locality.

\begin{figure}
    \centering
    \includegraphics[width=1\linewidth]{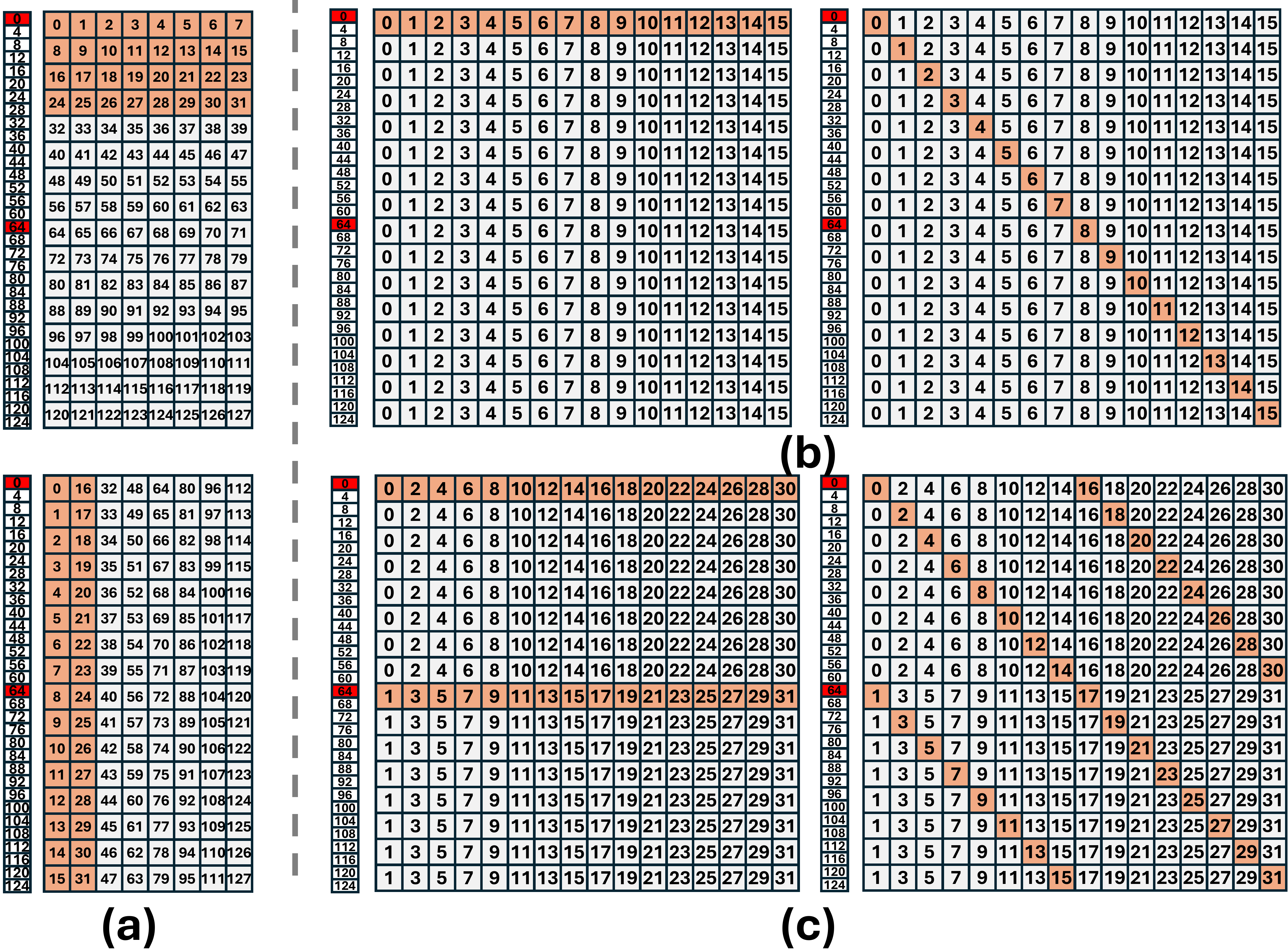}
    \caption{Shared Memory Access: FFT-CGEMM}
    \label{fig:shmem_fft_cgemm}
\end{figure}
While aligning the FFT and CGEMM workflows enables dataflow fusion, efficiently \textbf{forwarding FFT output into shared memory} in a layout compatible with CGEMM is non-trivial. In TurboFNO, achieving \textbf{100\% shared memory bank utilization} during the FFT-to-CGEMM forwarding process requires each thread to write \textbf{consecutive elements from the same signal} into shared memory. However, such a layout conflicts with the conventional FFT-internal access patterns, which are typically optimized to avoid bank conflicts during butterfly computation.

To address this, TurboFNO applies a customized \textbf{thread-level data layout and swizzling strategy} that resolves this tension. Figure~\ref{fig:shmem_fft_cgemm} illustrates our approach. On the left side of each subfigure, memory banks are shown as vertical strips (32 banks, 4 bytes each), while each small square on the right represents a single-precision complex number (8 bytes, occupying two banks). The number inside each square indicates the thread ID responsible for that data element during FFT computation.

In Figure~\ref{fig:shmem_fft_cgemm}(a, top), we show the layout used by prior works such as VkFFT, where \textbf{consecutive threads write to different signals at the same offset}. This avoids bank conflicts during FFT internal stages, but when forwarding the result to CGEMM via shared memory, it leads to severe conflicts. As shown in the orange highlights, thread groups 0--7, 8--15, etc., collide on the same set of banks, resulting in only \textbf{25\% warp-level bank utilization}. Worse, this conflict pattern cannot be removed by simple swizzling, since thread-to-bank mapping is static; resolving it would require memory padding, which wastes shared memory and hurts performance.

In contrast, TurboFNO adopts a layout where \textbf{consecutive threads write consecutive elements from the same FFT signal}, allowing CGEMM to consume the data in a column-major, bank-aligned fashion. This results in \textbf{100\% bank utilization} for FFT-to-CGEMM data forwarding, as shown in Figure~\ref{fig:shmem_fft_cgemm}(a, bottom).

However, this layout introduces a new problem during the final stages of FFT computation, when threads write their results back to shared memory. Figures~\ref{fig:shmem_fft_cgemm}(b) and (c) illustrate this challenge for the two most common FFT workloads: 16-point FFT and 8-point FFT, respectively. In Figure~\ref{fig:shmem_fft_cgemm}(b, left), all threads (0--15) attempt to write to the same few banks, causing severe conflicts and reducing utilization to just \textbf{6.25\%} (2 out of 32 banks active). To overcome this, we observe that each thread holds a unique subset of the output, and together they cover the full address space. By applying an \textbf{offset proportional to the thread ID} when writing to shared memory (i.e., \texttt{addr += tid}), we can fully eliminate the conflict and restore \textbf{100\% bank utilization}, as shown in Figure~\ref{fig:shmem_fft_cgemm}(b, right). Figure~\ref{fig:shmem_fft_cgemm}(c) shows the same strategy applied to the 8-point FFT case. Since neighboring threads do not conflict (e.g., thread 0 and 1 access banks 0 and 64, respectively), a smaller offset such as \texttt{addr += tid / 2} is sufficient to avoid collisions. This two-level design—(1) column-major FFT output layout for CGEMM, and (2) thread swizzling within FFT—allows TurboFNO to achieve \textbf{bank-conflict-free access} both during FFT computation and while forwarding data to CGEMM, without introducing memory padding overhead.

\subsection{Fused GEMM-iFFT}

\begin{figure}
    \centering
    \includegraphics[width=1\linewidth]{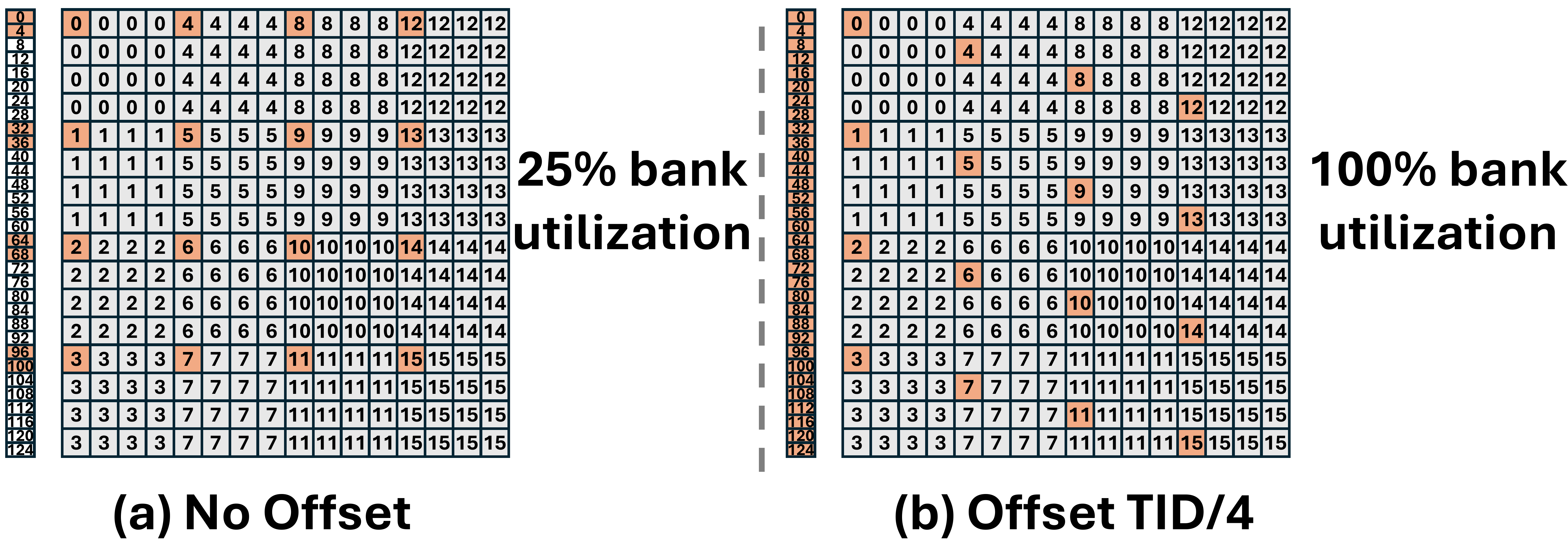}
    \caption{Shared Memory Access: CGEMM-iFFT}
    \label{fig:shmem_cgemm_ifft}
\end{figure}
Figure~\ref{fig:workflow_alignment}(f) illustrates how TurboFNO fuses the inverse FFT (iFFT) as an \textbf{epilogue} to the CGEMM kernel. Upon completion of CGEMM, the result matrix $C$ is written into shared memory along the $N$ direction. Crucially, TurboFNO only stores the subset of data required as iFFT input—matching the batch size used for the input slice in Figure~\ref{fig:workflow_alignment}(d). This reuse of batch slicing ensures memory efficiency and avoids unnecessary data movement. Each warp in TurboFNO is responsible for a \texttt{32$\times$16} output tile, with each thread within a warp handling a local \texttt{4$\times$4} region. However, this mapping naturally leads to \textbf{multiple threads targeting the same shared memory bank}, since the 32 threads in a warp are distributed over overlapping memory regions. Fortunately, each thread writes 4 rows (i.e., spans 4 memory banks), which opens an opportunity for \textbf{bank conflict elimination via swizzling}. Figure~\ref{fig:shmem_cgemm_ifft} provides a detailed view of this scenario. Each thread is responsible for writing a $4\times4$ tile of complex numbers into shared memory. Without swizzling, as shown in Figure~\ref{fig:shmem_cgemm_ifft}(a), threads such as 0, 4, 8, and 12 would simultaneously access memory banks 0 and 4, leading to high contention and poor utilization. To resolve this, TurboFNO employs a \textbf{swizzling strategy} depicted in Figure~\ref{fig:shmem_cgemm_ifft}(b). By adding an offset proportional to \texttt{threadIdx.x / 4} to each thread’s shared memory write address, we effectively stagger accesses to avoid overlapping banks. This layout achieves \textbf{100\% shared memory bank utilization} while preserving the expected data layout for the following iFFT stage. This design completes the full fusion path from FFT $\rightarrow$ CGEMM $\rightarrow$ iFFT, minimizing global memory traffic and ensuring high-throughput intra-kernel data reuse.

\subsection{Fused FFT-GEMM-iFFT}

By aligning the FFT and CGEMM workflows as shown in Figure~\ref{fig:workflow_alignment}, and resolving shared memory bank conflicts following the strategies depicted in Figures~\ref{fig:shmem_fft_cgemm} and~\ref{fig:shmem_cgemm_ifft}, we construct a high-performance fused FFT-CGEMM-iFFT kernel for TurboFNO.

\begin{figure}
    \centering
    \includegraphics[width=1\linewidth]{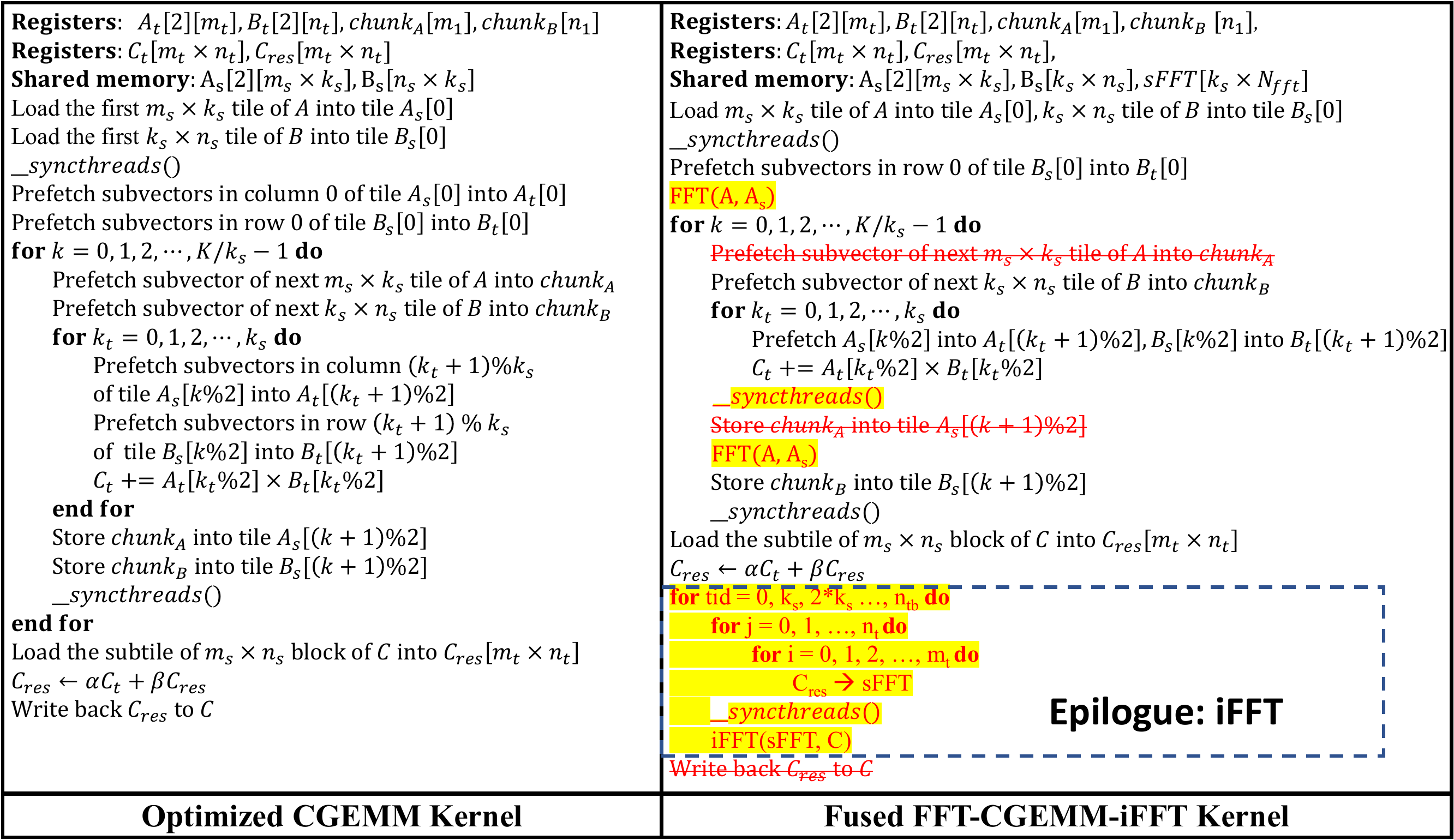}
    \caption{Pseudocode}
    \label{fig:pseudocode}
\end{figure}

Figure~\ref{fig:pseudocode} compares the fused kernel with the original CGEMM workflow. In this pseudocode visualization, the deleted operations are marked with red strikethrough lines, while the newly added modules are highlighted in yellow. As shown, all memory operations related to the operand $A$ fragment have been removed, including global memory prefetching and storing $A$ into shared memory. Instead, $A$ is treated as the input to the FFT module, and forwarded directly into TurboFNO's built-in FFT. The first FFT is executed once before the $k$-loop begins, mimicking the role of prefetching. Inside the loop, \texttt{FFT($A$)} replaces the original step where $A$ fragments were stored to shared memory. After completing the $k$-loop in CGEMM, the usual writeback of the output matrix $C$ to global memory is omitted. Instead, $C$ is forwarded to shared memory and used as the input to the \texttt{iFFT}, which is invoked once the corresponding shared memory tile becomes available. This iFFT operation is performed iteratively over the thread block’s $N$ dimension tiles, completing the end-to-end fusion path within a single kernel. This design eliminates all intermediate global memory writes and reads between FFT, CGEMM, and iFFT stages, significantly improving locality and reducing memory traffic.
\begin{table}[ht] \centering
\caption{CGEMM and FFT kernel parameter setup.}
\begin{tabular}{l
    S[table-format=3] 
    S[table-format=3] 
    S[table-format=3] 
    S[table-format=3] 
    S[table-format=3] 
    S[table-format=3] 
    S[table-format=3]
    }
\toprule
                   & {$m_{tb}$}       & {$n_{tb}$}       & {$k_{tb}$}  & {$m_w$} & {$n_w$} & {$m_t$} & {$n_t$} \\ 
\midrule
CGEMM              & {$32$}          & {$32$}             & {$8$}    & {$32$}  & {$16$}    & {$4$}  & {$4$} \\ 
\midrule
          & {$N_1$} &    {$N_2$} & {$n_{1}$}      & {$n_2$}       &  {$bs$}  \\ 
\midrule
FFT & 128 & 256 & 8 & 16 & 8\\
\bottomrule
\end{tabular}
\label{tab:kernel_size}
\vspace{-4mm}
\end{table}
Table~\ref{tab:kernel_size} summarizes the kernel configuration parameters used in our CGEMM and FFT implementations. For the FFT kernel, $N_i$ denotes the signal length handled at the threadblock level, while $n_i$ represents the per-thread FFT size corresponding to $N_i$. The parameter \texttt{bs} indicates the number of signals of length $N_i$ processed by each threadblock—i.e., the FFT batch size per block. This batch size is designed to align with the CGEMM’s threadblock-level tile width, \texttt{k\_tb}, to ensure dataflow compatibility. In our configuration, we set \texttt{bs} = 8 to match the CGEMM blocking dimensions.

\section{Performance Evaluation}
\label{sec:results}
\begin{figure}[]
    \centering
    \includegraphics[width=\linewidth]{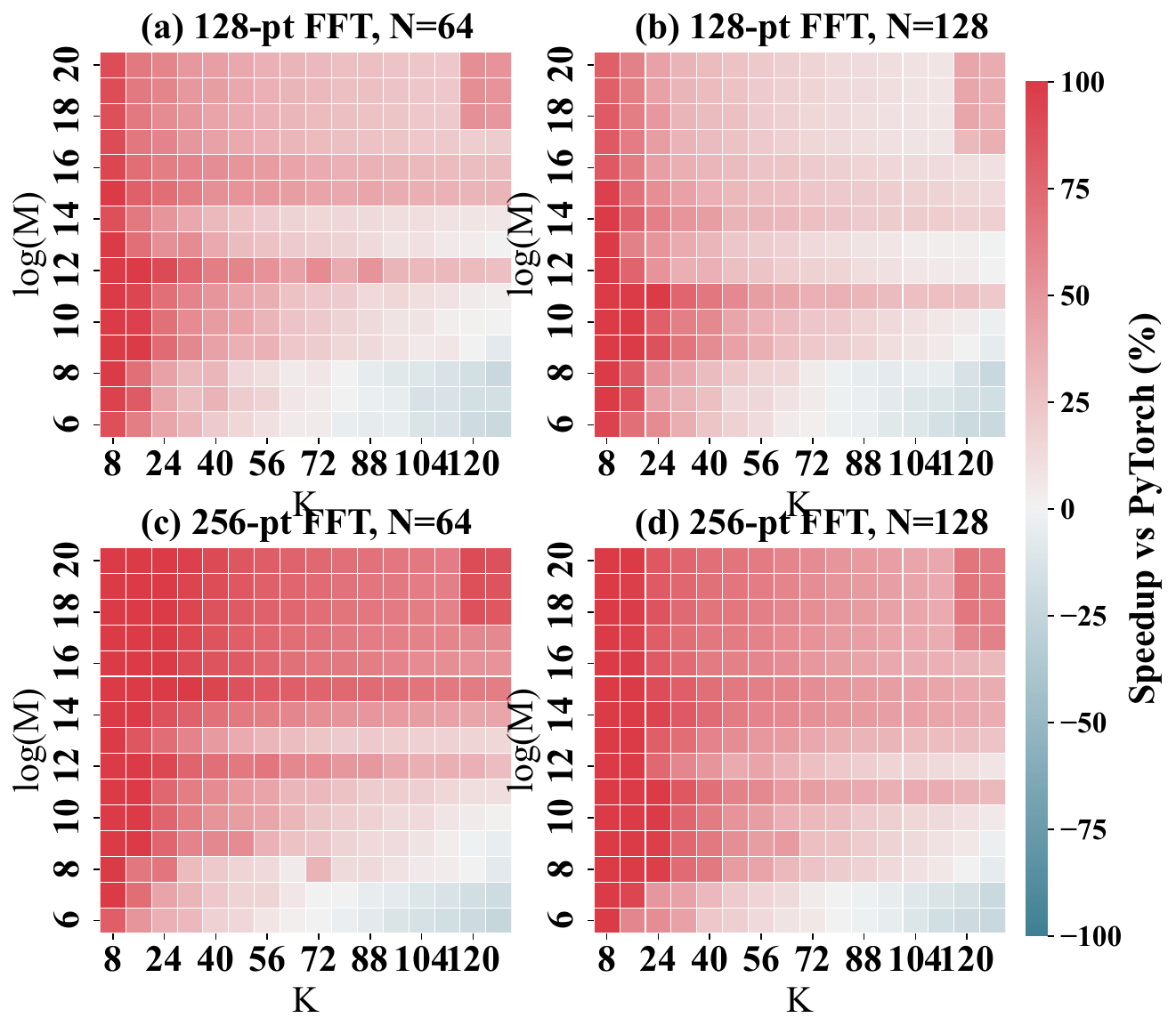}
    \caption{1D TurboFNO vs PyTorch}
    \label{fig:1D_heatmap}
\end{figure}

We evaluate TurboFNO on NVIDIA 40GB A100-PCIE GPU. The A100 GPU is connected to a node with one 64-core AMD EPYC 7763 CPU with a boost frequency of 3.5 GHz. We compile programs using CUDA $\mathtt{12.4}$. We present the performance evaluations of step-wise CGEMM and single precision FFT optimizations, FFT truncation, FFT pruning, fused FFT-CGEMM, fused-CGEMM-iFFT, and fused FFT-CGEMM-iFFT. We compare our TurboFNO with the state-of-the-art Fourier neural operator implementation in PyTorch, which implemented with NVIDIA closed-source library cuBLAS, cuFFT and PyTorch built-in memory kernel. We also compare our fused FFT-CGEMM-iFFT with cuBLAS CGEMM and cuFFT C2C with CUDA $\mathtt{12.4}$. To ensure a fair comparison with PyTorch, we manually implemented a CUDA C baseline that directly invokes \texttt{cuFFT} and \texttt{cuBLAS}, replicating the computational behavior of the PyTorch implementation. This approach avoids Python-level overhead and isolates kernel-level performance differences. The reported performance data are averaged over tens of runs to minimize fluctuations. For better understandability, Table \ref{tab:fft_overhead_comparison} lists the method and the comparison base in 1D FFT evaluation ( Figure \ref{fig:1D_bar_A}-\ref{fig:1D_heatmap}) and 2D FFT evaluation (Figure \ref{fig:2D_bar_A}-\ref{fig:2D_heatmap}). Figure \ref{fig:1D_bar_A} $\rightarrow$ Figure \ref{fig:1D_bar_B}$\rightarrow$ Figure \ref{fig:1D_bar_C}$\rightarrow$ Figure \ref{fig:1D_bar_D}$\rightarrow$ Figure \ref{fig:1D_heatmap} presents the stepwise optimization of our built-in FFT Pruning, FFT Truncation, FFT Zero-Padding, and fused FFT-CGEMM-iFFT in TurboFNO.

\begin{table}[ht]
\centering
\caption{Method and Comparison Base in Evaluation.}
\label{tab:fft_overhead_comparison}
\begin{tabular}{|c|c|l|l|}
\hline
\textbf{Id}& Figure & \textbf{TurboFNO Optimization} & \textbf{ Base} \\ \hline
A & Fig. \ref{fig:1D_bar_A},\ref{fig:2D_bar_A} & FFT Pruning, Truncation & PyTorch\\ \hline
B & Fig. \ref{fig:1D_bar_B},\ref{fig:2D_bar_B} & Fused FFT-CGEMM & PyTorch, A  \\ \hline
C & Fig. \ref{fig:1D_bar_C},\ref{fig:2D_bar_C} & Fused CGEMM-iFFT & PyTorch, A, B \\ \hline
D & Fig. \ref{fig:1D_bar_D},\ref{fig:2D_bar_D} & Fused FFT-CGEMM-iFFT & PyTorch, A, B, C \\ \hline
E & Fig. \ref{fig:1D_heatmap},\ref{fig:2D_heatmap} & TurboFNO: A+B+C+D & PyTorch \\ \hline
\end{tabular}
\end{table}
\subsection{1D FNO Evaluation}
\subsubsection{FFT Pruning, Truncation and Zero Padding}
\begin{figure}[]
    \centering
    \includegraphics[width=\linewidth]{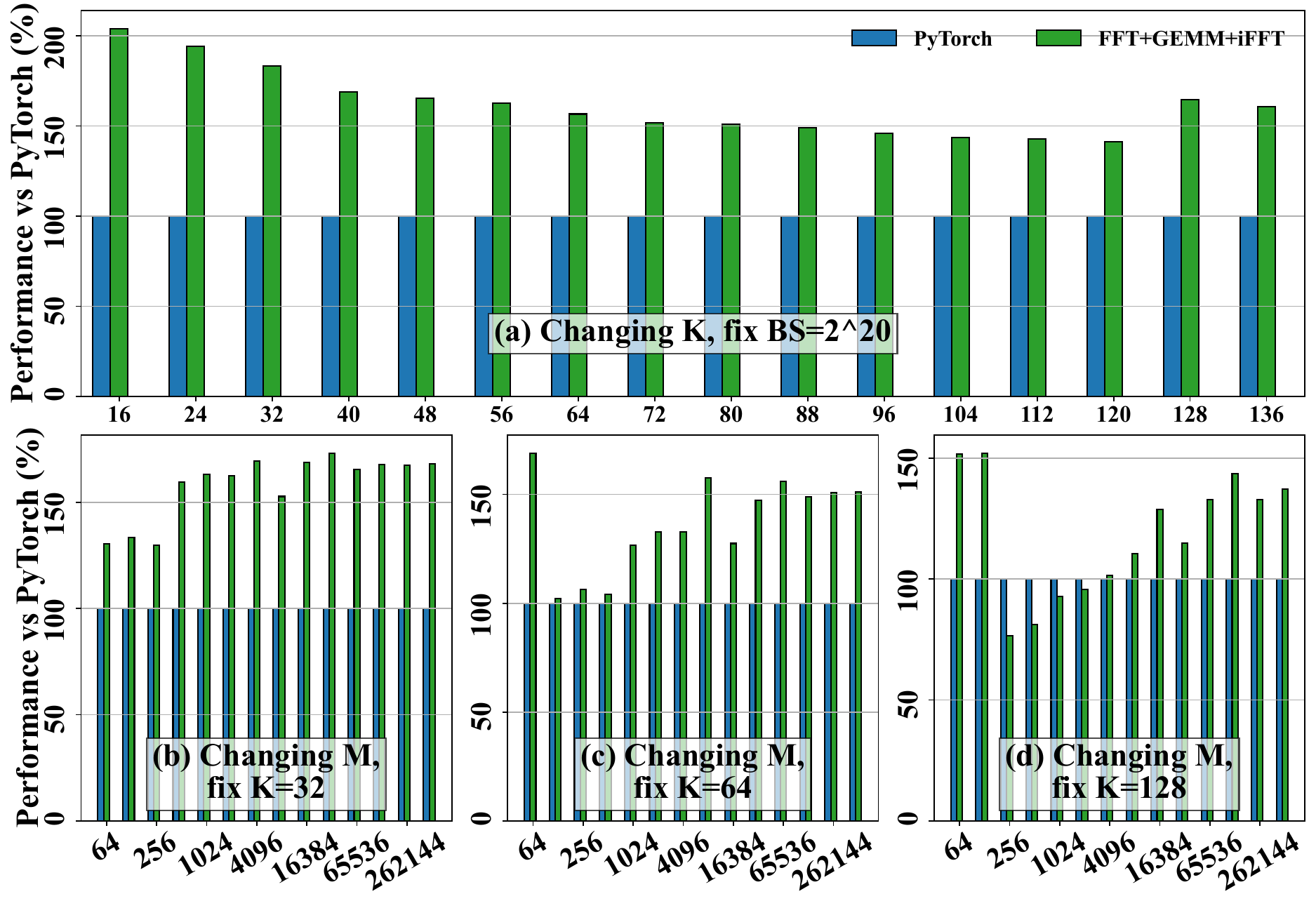}
    \caption{1D FFT Optimization}
    \label{fig:1D_bar_A}
\end{figure}

Figure~\ref{fig:1D_bar_A} demonstrates that our optimized FFT-CGEMM-iFFT workflow achieves up to 100\% speedup over PyTorch, with an average speedup of 50\%. This improvement is primarily attributed to our FFT pruning strategy, which reduces computation by 25\%--67.5\%, as well as to FFT truncation and zero-padding, which reduce global memory writes (in FFT) and reads (in iFFT) by a factor of \texttt{Filter\_size / Input\_size}.

As shown in Figure~\ref{fig:1D_bar_A}(a), when the hidden dimension $K$ is small (e.g., 16 to 32), our optimized FFT achieves a 70\%--100\% speedup compared to PyTorch. However, as $K$ increases, the speedup stabilizes around 50\%. This behavior is due to our FFT implementation modifying the data access pattern from $(X, Y)$ to $(Y, \text{HiddenDim})$ to align with CGEMM’s dataflow. While this benefits fused execution, it reduces L1 cache locality across thread blocks. Furthermore, the workload per thread block shifts from processing a contiguous chunk of size \texttt{threadblock\_bs} $\times$ \texttt{signal\_length} (where \texttt{signal\_length} is the FFT length and \texttt{threadblock\_bs} is the number of FFTs handled by each thread block) to iteratively traversing the hidden dimension like a $k$-loop in GEMM. Although this structure prepares for fusion, it causes minor performance degradation.

Figures~\ref{fig:1D_bar_A}(b)--(d) show performance under fixed $K = 32, 64, 128$ while varying $BS$ (i.e., the batch size). In all three cases, we observe that the \textbf{speedup ratio increases with $BS$}. This is because as the problem size grows, the global memory savings from TurboFNO's FFT increasingly outweigh the negative effects of the altered dataflow, ultimately becoming the dominant factor in performance.

\subsubsection{Kernel Fusion of FFT and CGEMM}
\begin{figure}[ht]
    \centering
    \includegraphics[width=\linewidth]{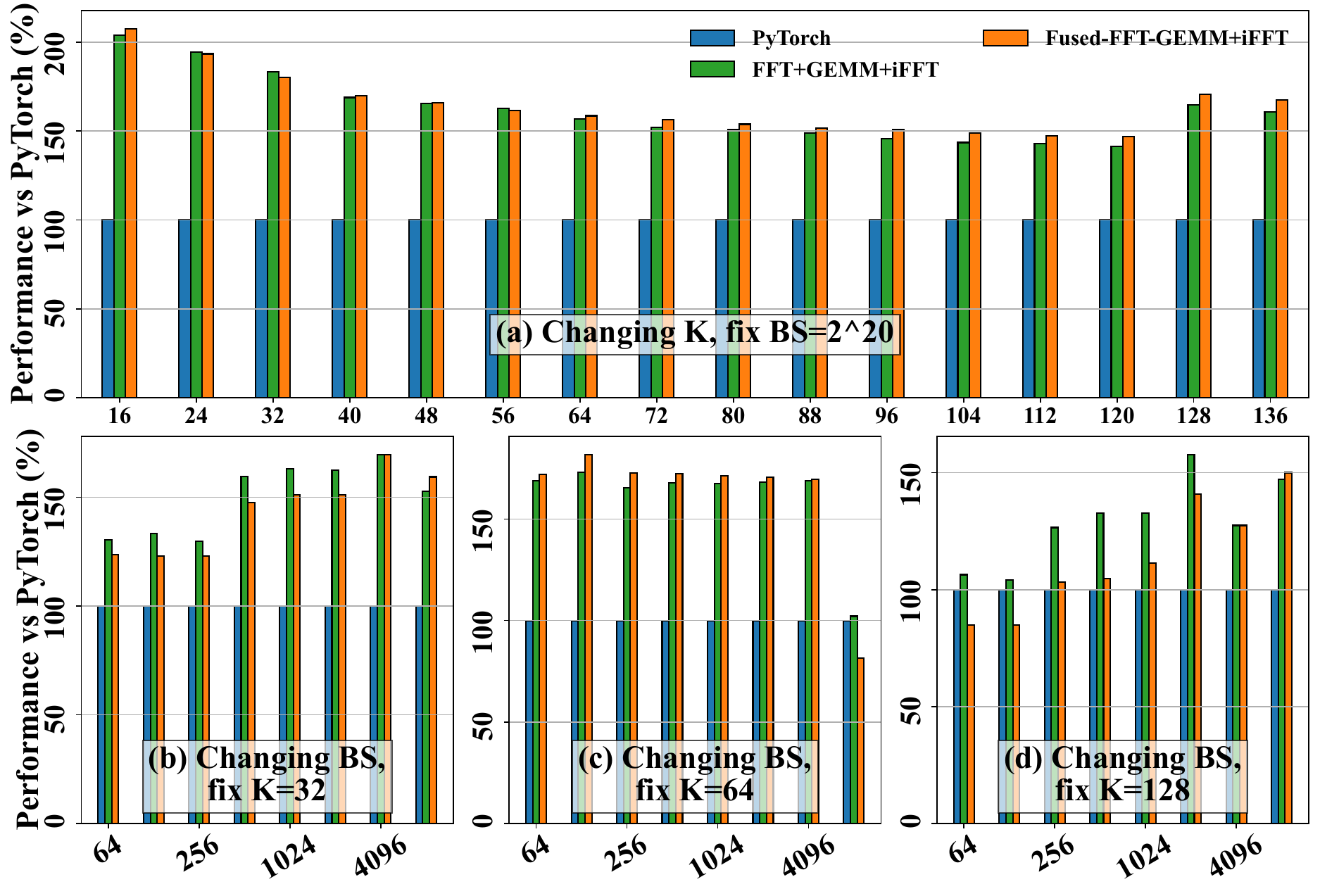}
    \caption{1D Fused FFT-CGEMM}
    \label{fig:1D_bar_B}
\end{figure}

Figure~\ref{fig:1D_bar_B} evaluates the effect of fusing the forward FFT and CGEMM kernels on top of our FFT optimizations. We compare the fused FFT-CGEMM kernel with both PyTorch and the non-fused FFT+CGEMM+iFFT baseline, which applies only our FFT optimization without kernel fusion. Compared to PyTorch, the fused version achieves a 50\%--100\% speedup, while the improvement over the non-fused FFT-optimized workflow is only 3\%--5\%. Furthermore, as shown in Figures~\ref{fig:1D_bar_B}(b)--(d), increasing the hidden dimension $K$ from 32 to 128 leads to a gradual decline in the benefits of kernel fusion. For large hidden dimensions ($K \geq 128$), fusion may even degrade performance. This indicates that, in such cases, the memory savings from kernel fusion are insufficient to compensate for the performance loss introduced by modifying the FFT and CGEMM workflow. However, as shown in Figure~\ref{fig:1D_bar_B}(a), when the input size $BS$ is large (e.g., $BS = 2^{20}$), the benefits of kernel fusion remain consistently positive, highlighting its effectiveness in large-scale settings.

\subsubsection{Kernel Fusion of CGEMM and iFFT}
\begin{figure}[ht]
    \centering
    \includegraphics[width=\linewidth]{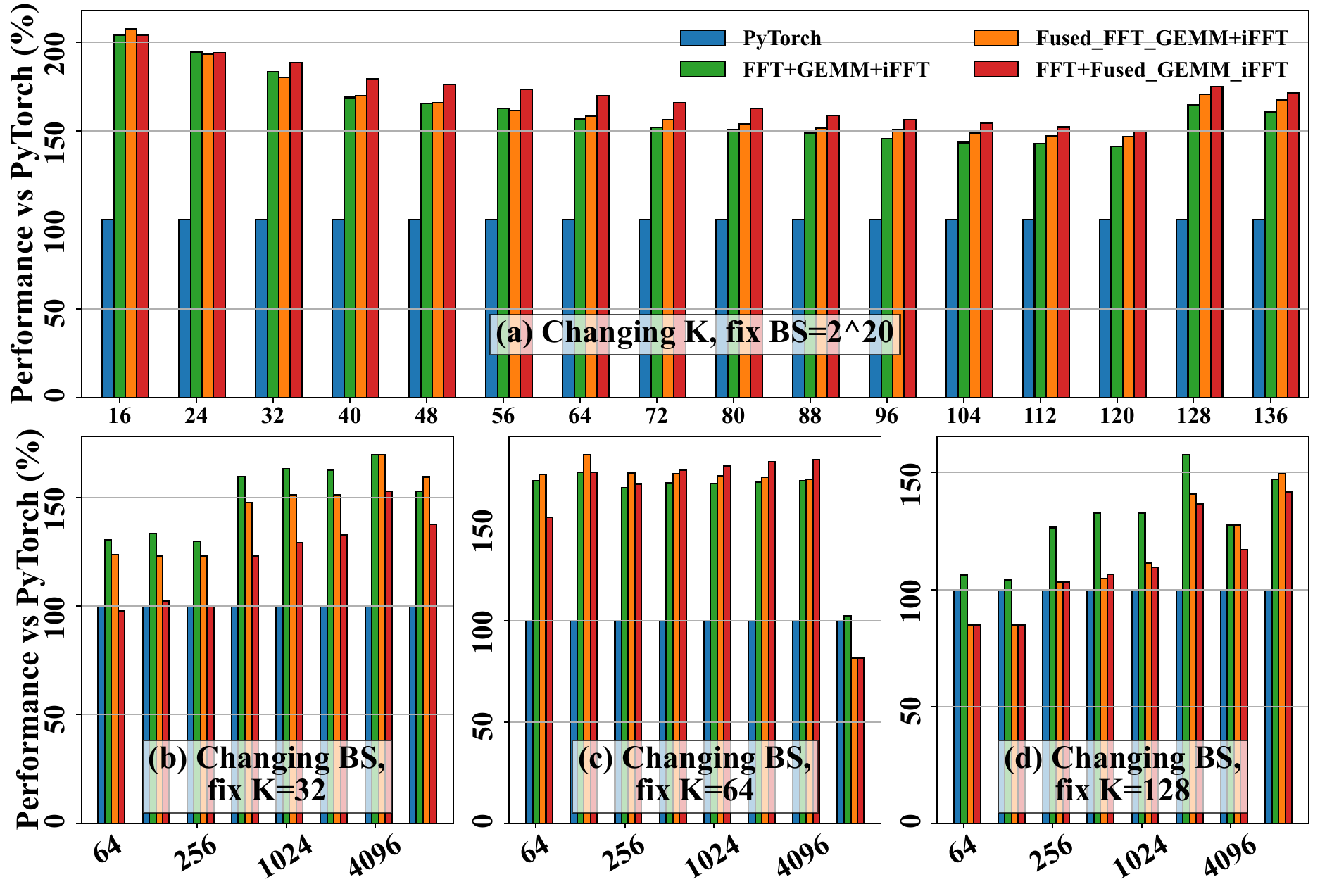}
    \caption{1D Fused CGEMM-iFFT}
    \label{fig:1D_bar_C}
\end{figure}

Figure~\ref{fig:1D_bar_C} shows that the fusion of CGEMM and iFFT epilogue, combined with our FFT optimizations, further improves TurboFNO’s performance. As illustrated in Figure~\ref{fig:1D_bar_C}(a), this gain is largely due to our carefully designed shared memory access pattern for writing \texttt{C\_result}, which achieves 100\% bank conflict-free utilization. Compared to both the FFT-only optimization and the forward FFT+CGEMM fusion, the fused CGEMM-iFFT kernel achieves at least a 50\% speedup over PyTorch across all problem sizes shown.

By contrast, the previous two optimization strategies—FFT-only and forward FFT+CGEMM—exhibit slightly lower speedups than 50\% when $K$ is around 104--120. Notably, at $K = 32$ and $K = 128$, kernel fusion of CGEMM and iFFT introduces a 10\%--30\% performance degradation compared to FFT-only optimization. However, as seen in Figure~\ref{fig:1D_bar_C}(c), CGEMM-iFFT fusion still offers a 5\% improvement over FFT-only in certain configurations. This discrepancy is likely due to the specific CGEMM kernel parameters used—$M_{tb}$ = 64, $N_{tb}$ = 128, $K_{tb}$ = 8, $M_{w}$ = 32, $N_{w}$ = 16, $M_{t}$ = 4, and $N_{t}$ = 4—which may not be optimal for these particular hidden dimensions, or may adversely affect cache utilization at those values of $K$.

\subsubsection{Fused FFT-CGEMM-iFFT}
\begin{figure}[]
    \centering
    \includegraphics[width=\linewidth]{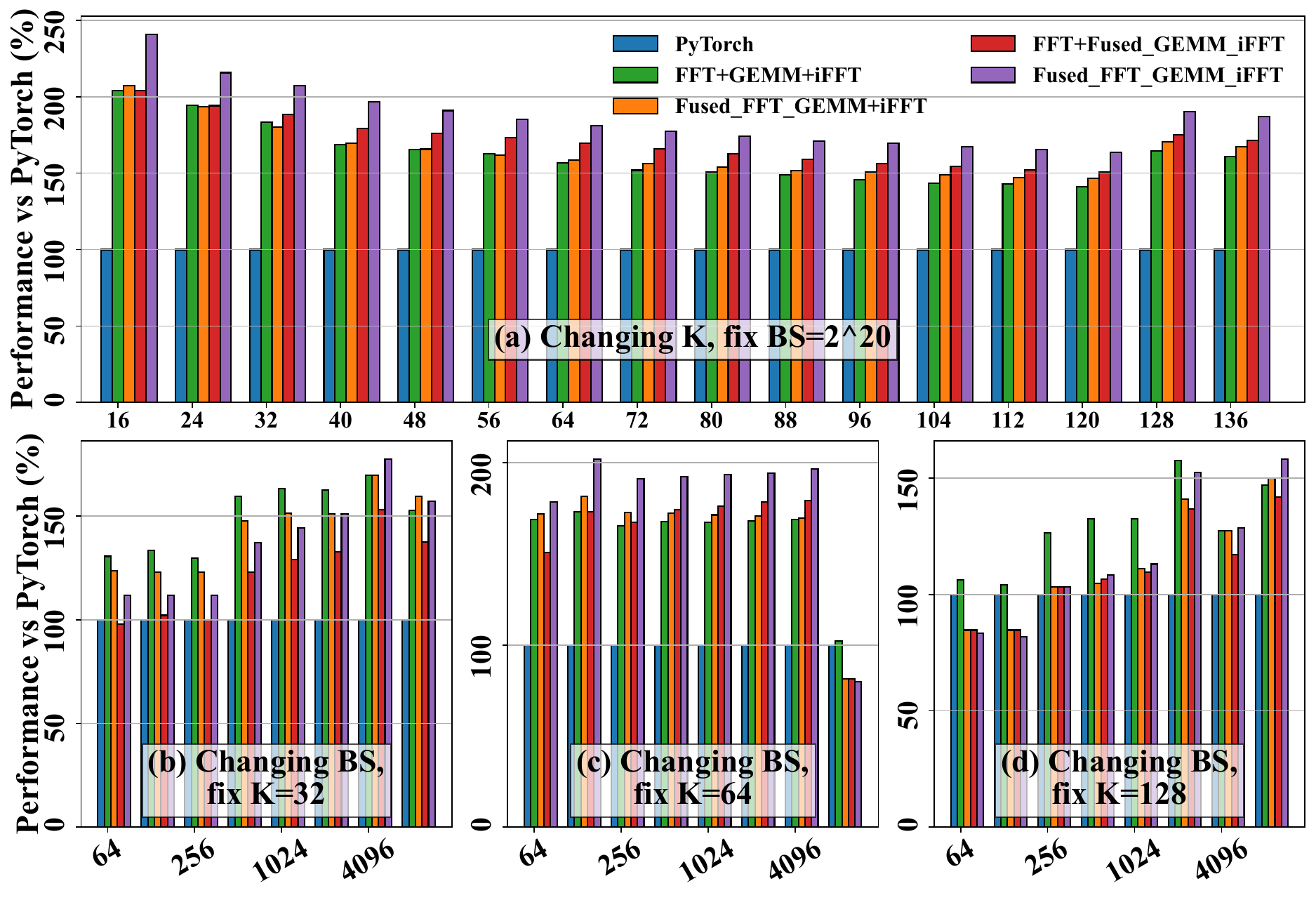}
    \caption{1D Fused FFT-CGEMM-iFFT}
    \label{fig:1D_bar_D}
\end{figure}

Figure~\ref{fig:1D_bar_D} demonstrates that the fully fused FFT-CGEMM-iFFT kernel in TurboFNO substantially outperforms the previously discussed partial fusion strategies, namely FFT-CGEMM and CGEMM-iFFT. As shown in Figures~\ref{fig:1D_bar_D}(a) and (c), our fully fused design achieves up to 150\% speedup over PyTorch, and delivers an additional 10\%--20\% improvement over prior TurboFNO optimization stages.

These results confirm that memory transaction reduction is the primary performance bottleneck in Fourier Neural Operators. In Figure~\ref{fig:1D_bar_D}(a), for large problem sizes, our fully fused pipeline demonstrates that despite deviating from the standard FFT workflow and introducing an extra \texttt{\_\_syncthreads()} in CGEMM, performance still improves significantly. This is attributed to two key factors: (1) the bank-conflict-free shared memory write of FFT results that conforms to CGEMM's access pattern, and (2) the bank-conflict-free layout of CGEMM output results to match the iFFT input access pattern. These careful shared memory designs enable the fully fused FFT-CGEMM-iFFT kernel to deliver further gains.

Additionally, the results suggest that the FFT and CGEMM workflows still contain computational gaps that allow fusion to improve performance beyond simple kernel chaining.

However, Figures~\ref{fig:1D_bar_D}(b) and (d) show that for certain problem sizes, the performance of the fully fused FFT-CGEMM-iFFT kernel slightly degrades compared to the fused FFT-CGEMM or FFT-only optimization. This is mainly due to the underperformance of the CGEMM-iFFT fusion in these specific cases. Although the fused FFT-CGEMM stage partially hides the iFFT epilogue overhead, it cannot eliminate it entirely, especially in configurations without a well-optimized epilogue.

To summarize these findings, Figure~\ref{fig:1D_heatmap} presents a comprehensive comparison between the best-performing version of TurboFNO with all optimizations and PyTorch across a wide range of configurations.

\subsubsection{TurboFNO: Full Optimization Results}
Figure~\ref{fig:1D_bar_D} presents the best-performing configuration among all TurboFNO strategies. Compared to PyTorch, it achieves an average speedup of 44\% and a maximum speedup of up to 250\%. Subfigures~(a), (b), (c), and (d) illustrate different combinations of FFT sizes (128 and 256) and signal truncation dimensions (64 and 128), respectively.

In each subplot, the $x$-axis represents the hidden dimension $K$, while the $y$-axis denotes the logarithm of the product of batch size and input dimension $X$. The color intensity indicates speedup—deeper red corresponds to higher speedup, whereas blue indicates cases where TurboFNO is slower than PyTorch.

As observed, slowdowns (i.e., blue regions) only appear in the \textbf{lower-left region} of each heatmap, corresponding to small batch sizes and large $K$. In these cases, TurboFNO assigns one thread block to process along the $(Y, K)$ dimensions to match the CGEMM and FFT workflows, resulting in suboptimal SM utilization.

Overall, this figure demonstrates that TurboFNO effectively improves the performance of Fourier Neural Operators compared to the state-of-the-art PyTorch implementation (built upon cuFFT and cuBLAS), especially in large-scale scenarios.

\subsection{2D FNO Evaluation}

\subsubsection{2D FFT Pruning, Truncation, and Zero Padding}
\begin{figure}[]
    \centering
    \includegraphics[width=\linewidth]{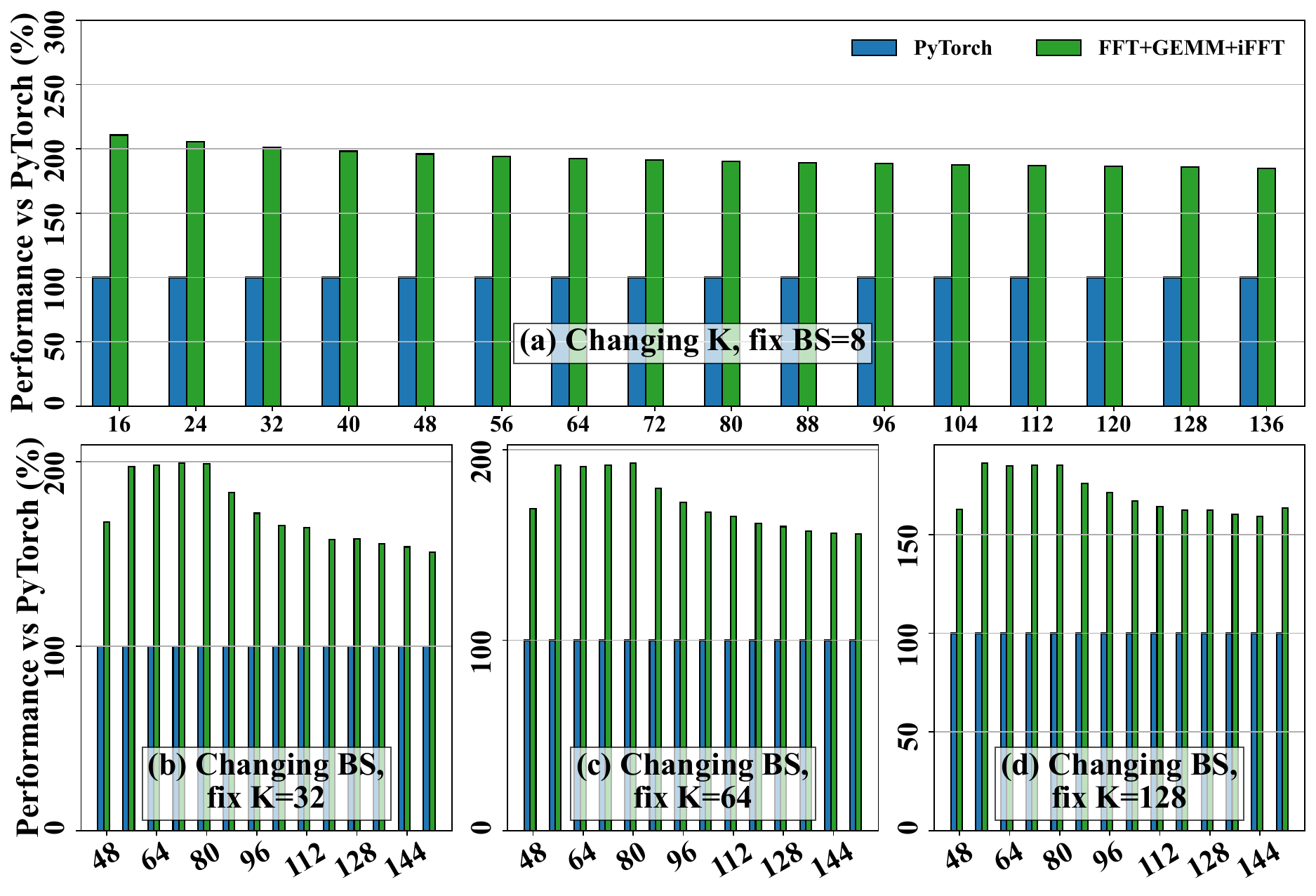}
    \caption{2D FFT Optimization}
    \label{fig:2D_bar_A}
\end{figure}

Figure~\ref{fig:2D_bar_A} demonstrates that our optimized FFT-CGEMM-iFFT workflow consistently outperforms PyTorch in the 2D setting, achieving an average speedup above 50\% and a maximum speedup up to 100\%. Similar to the 1D case, the primary contributors to this performance gain are our FFT pruning strategy—reducing computation by 25\%--67.5\%—and the use of FFT truncation and zero-padding, which reduce global memory writes (in the first FFT) and reads (in the second iFFT) by a factor of \texttt{Filter\_size / Input\_size}.

Compared to the 1D evaluation, where speedups under small problem sizes (e.g., $BS < 1024$) only reached around 30\%, the 2D case exhibits much more stable and higher speedups even for small-scale problems. For instance, in Figure~\ref{fig:2D_bar_A}(b), the leftmost region (small $BS$ and moderate $K$) shows speedups close to 100\%, underscoring the greater efficiency of our 2D design. This improvement reflects a shift in the performance bottleneck: in the 2D FFT pipeline, the combination of (1) reduced computation and memory reads from the first FFT due to pruning and truncation, and (2) reduced memory reads in the second FFT, has become the dominant factor. As a result, the TurboFNO gains in the 2D setting are not only larger but also more robust across different problem sizes.

\subsubsection{Fused FFT-CGEMM}
\begin{figure}[]
    \centering
    \includegraphics[width=\linewidth]{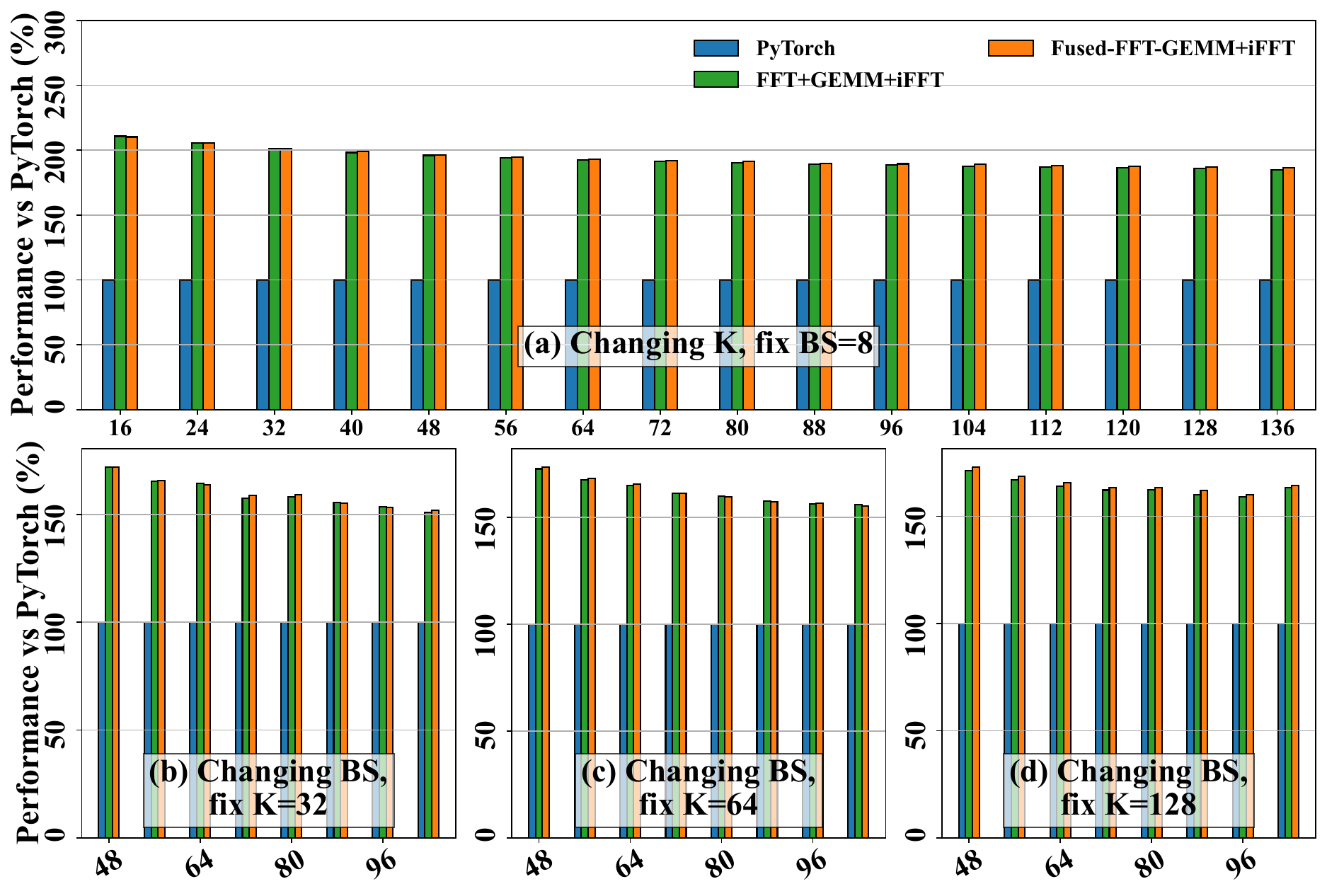}
    \caption{2D Fused FFT-CGEMM}
    \label{fig:2D_bar_B}
\end{figure}

Figure~\ref{fig:2D_bar_B} presents the performance of enabling fused FFT-CGEMM in the 2D TurboFNO pipeline. As shown, the improvement from fusion is relatively limited, yielding only about \textbf{1\%--2\%} speedup across the evaluated test set. This is primarily because the inclusion of the first FFT in the 2D setting significantly increases the global memory read/write overhead on the FFT side. As a result, this memory bottleneck dominates the overall execution time and masks the benefits of low-overhead fusion in the FFT-CGEMM-iFFT workflow.

\subsubsection{Fused CGEMM-iFFT}
\begin{figure}[]
    \centering
    \includegraphics[width=\linewidth]{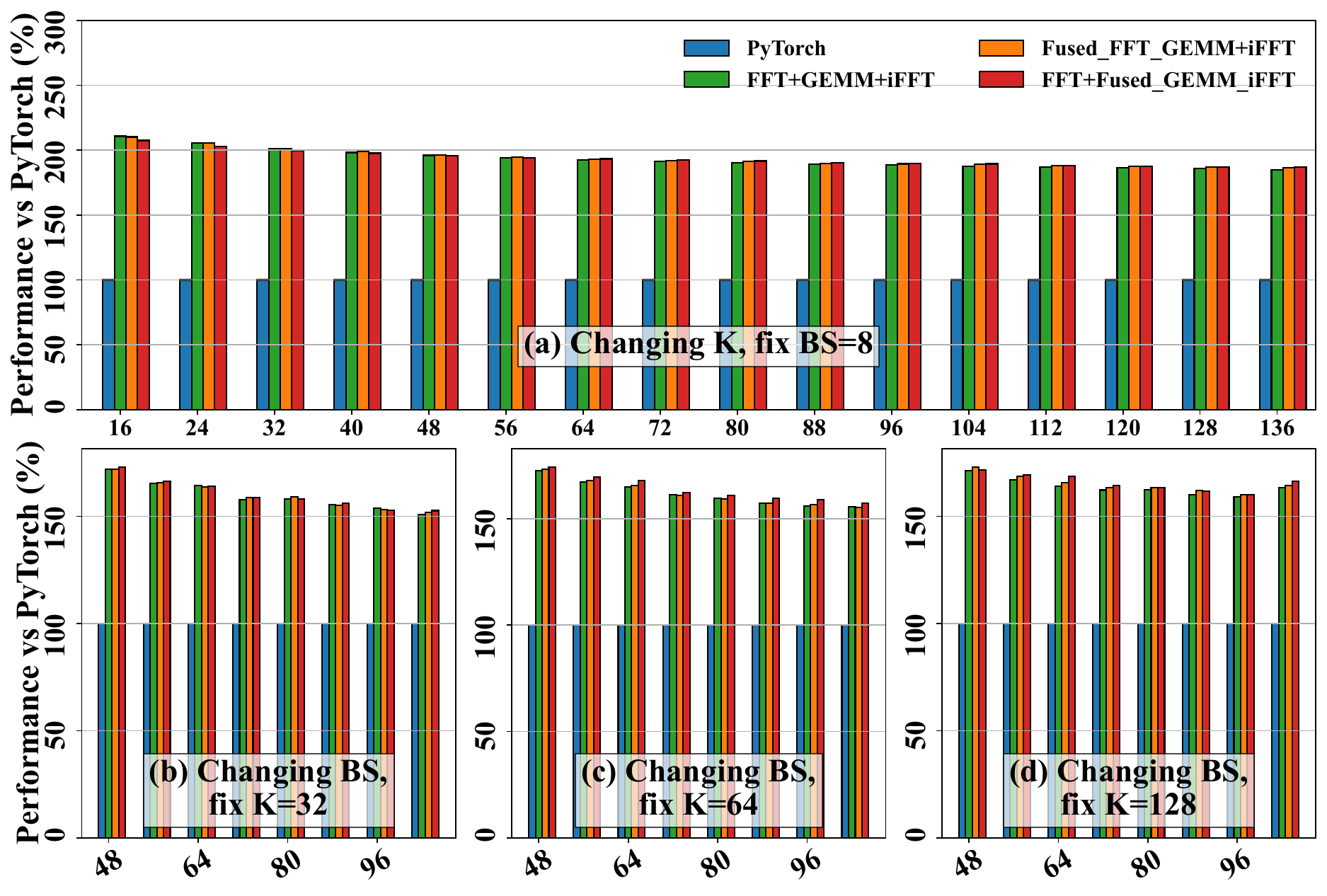}
    \caption{2D Fused CGEMM-iFFT}
    \label{fig:2D_bar_C}
\end{figure}

Figure~\ref{fig:2D_bar_C} illustrates the performance impact of fusing the CGEMM and iFFT kernels in the 2D TurboFNO pipeline. Compared to both the fused FFT-CGEMM and the FFT-only optimization baselines, our fused CGEMM-iFFT implementation still maintains a \textbf{50\%--100\% speedup over PyTorch}. While the additional gains from fusion over FFT-only optimization appear minimal in Figure~\ref{fig:2D_bar_C}(a), subfigures~(b), (c), and (d) show consistent improvements of \textbf{1\%--3\%}. The relatively modest gains can be attributed to the fact that in 2D FNO, the first-stage FFT introduces substantial global memory overhead, which dominates execution time and diminishes the relative benefit of kernel fusion in the later stages.

\subsubsection{2D Fused FFT-CGEM-iFFT}
\begin{figure}[]
    \centering
    \includegraphics[width=\linewidth]{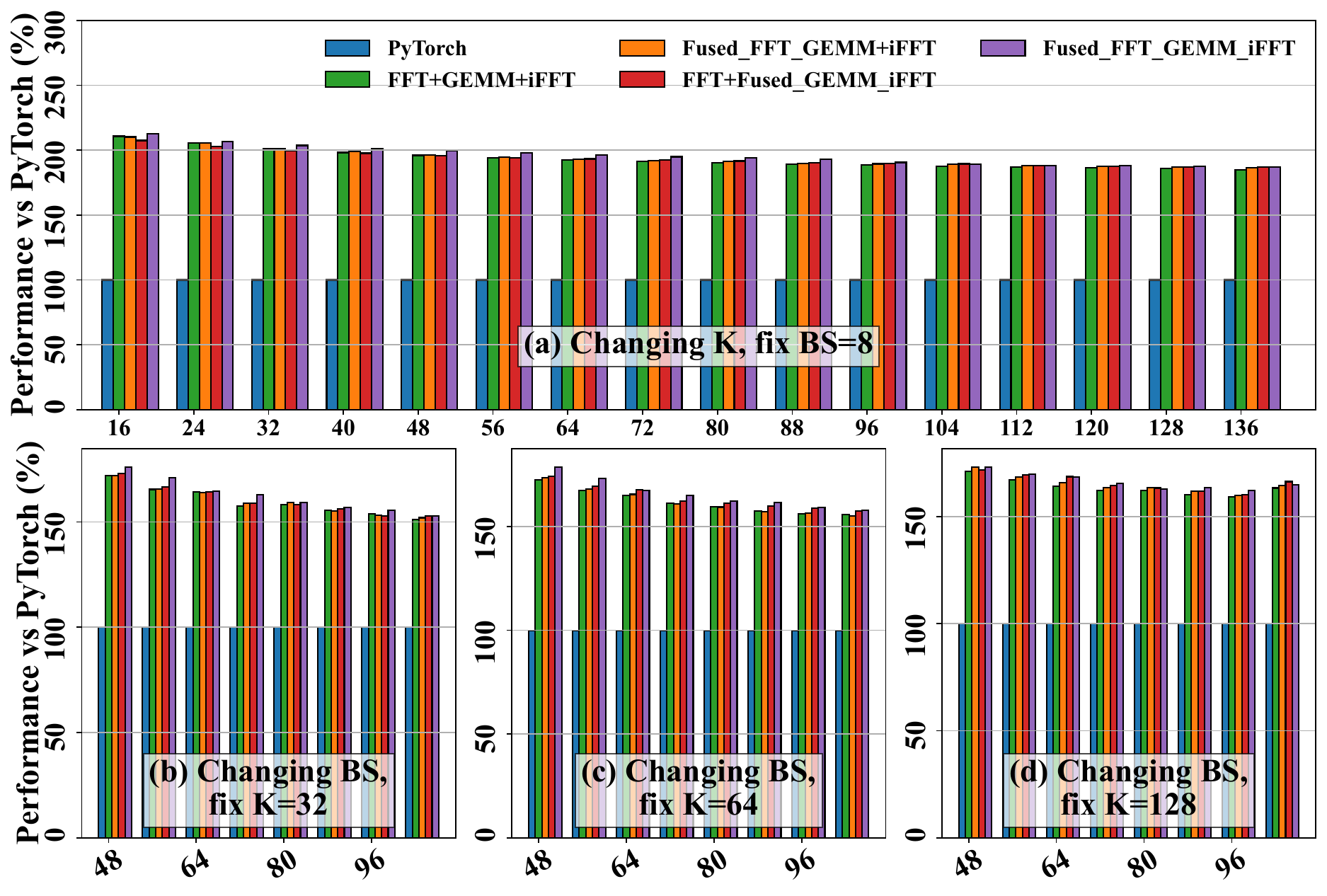}
    \caption{2D Fused FFT-CGEMM-iFFT}
    \label{fig:2D_bar_D}
\end{figure}

As shown in Figure~\ref{fig:2D_bar_D}, the fully fused FFT-CGEMM-iFFT kernel achieves a \textbf{50\%--105\% speedup} over PyTorch in the 2D setting. Notably, thanks to our carefully designed shared memory access pattern with \textbf{100\% memory bank utilization}, the additional overhead introduced by aligning the FFT workload layout to match CGEMM’s dataflow does not degrade performance. Moreover, Figures~\ref{fig:2D_bar_D}(b)--(d) demonstrate that the fully fused implementation consistently outperforms both the FFT-only and partially fused baselines (i.e., fused FFT-CGEMM or fused CGEMM-iFFT), yielding an additional \textbf{2\%--3\% performance improvement}. This highlights the effectiveness of our end-to-end fusion strategy, even in the presence of increased memory and synchronization complexity.

\subsubsection{2D TurboFNO: Full Optimization}
\begin{figure}[]
    \centering
    \includegraphics[width=\linewidth]{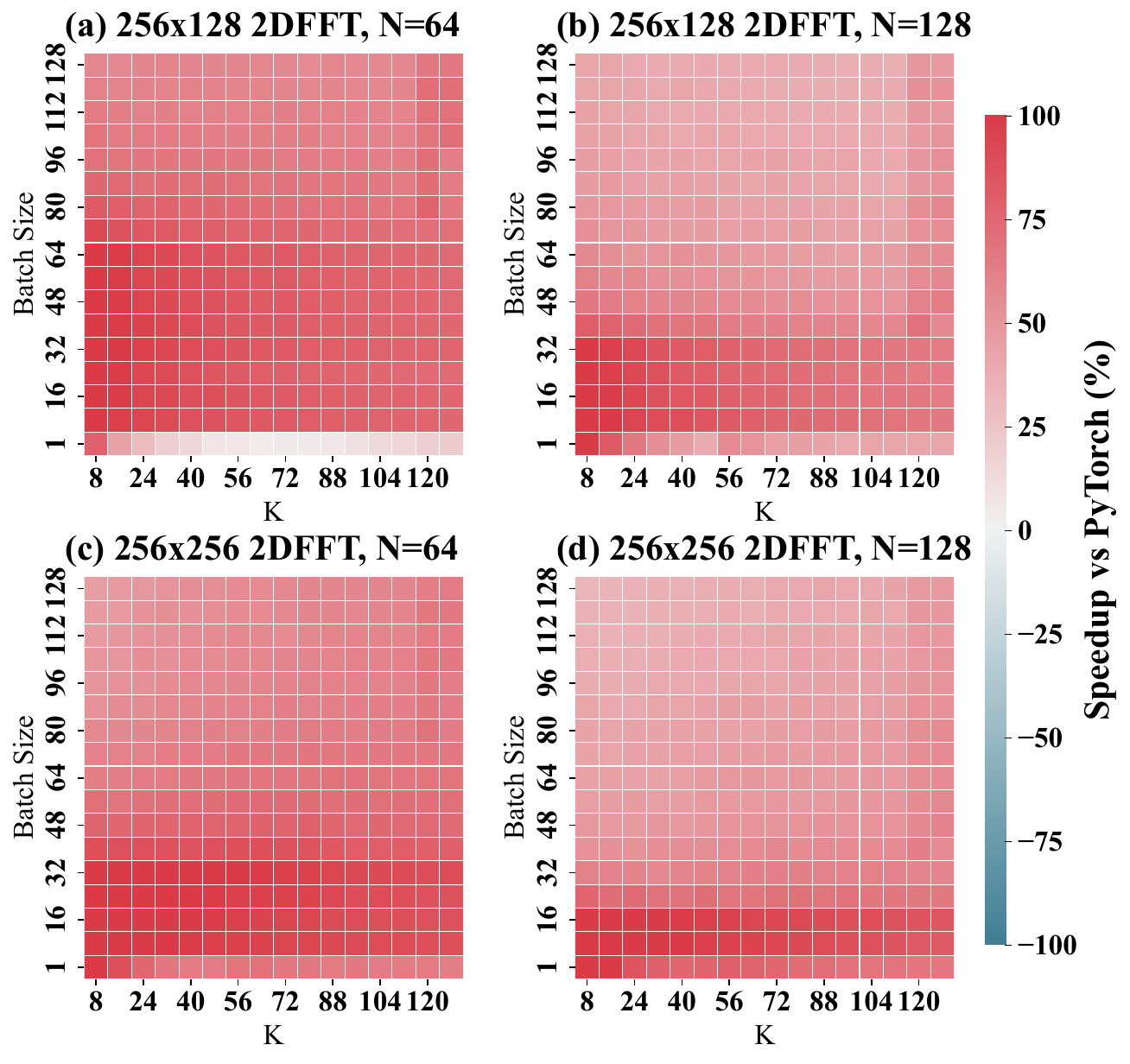}
    \caption{2D TurboFNO vs PyTorch}
    \label{fig:2D_heatmap}
\end{figure}

Figure~\ref{fig:2D_heatmap} summarizes the best-performing configuration of TurboFNO across all applied optimizations and compares it against PyTorch. On average, TurboFNO achieves a 67\% performance improvement, with a maximum speedup of 150\% observed across all evaluated problem sizes. These results highlight the effectiveness of TurboFNO’s comprehensive optimization pipeline, which includes FFT pruning, built-in truncation and zero-padding kernels, and fully fused FFT-CGEMM-iFFT execution. Together, these optimizations significantly reduce global memory traffic and computation overhead, leading to substantial acceleration over the state-of-the-art PyTorch implementation.

\section{Conclusion}

In this work, we propose \textbf{TurboFNO}, the first fully fused GPU kernel that integrates FFT, CGEMM, and iFFT for accelerating Fourier Neural Operators. To overcome the inefficiencies in conventional FNO implementations—such as excessive kernel launches and global memory traffic—we develop a series of architecture-aware optimizations. These include a custom FFT kernel with built-in frequency filtering and zero padding, a GEMM-compatible FFT variant that mimics $k$-loop behavior, and shared memory swizzling strategies that improve bank utilization from 25\% to 100\%. Additionally, we introduce a GPU-side FFT pruning mechanism that eliminates redundant computations for truncated frequencies. Our extensive evaluation shows that TurboFNO consistently outperforms the baseline PyTorch implementation across 1D and 2D problem settings. Specifically, TurboFNO achieves up to \textbf{150\% speedup} and maintains an average \textbf{67\% performance gain} across all tested configurations. Compared to prior strategies such as FFT-only optimization and partial kernel fusion, our fully fused FFT-CGEMM-iFFT pipeline delivers more stable and scalable performance, particularly in large-scale and 2D scenarios. These results confirm that memory transaction reduction—enabled by kernel fusion and fine-grained shared memory design—is the key to unleashing the full performance potential of FNO on modern GPUs.

\newpage
\bibliography{setting/Reference.bib}

\end{document}